\documentclass[twocolumn]{aastex631}
\usepackage{subfigure}

\shorttitle{Hierarchical structure in Maddalena}
\shortauthors{Shen et al.}
\graphicspath{{./}{figures/}}

\begin{document}

\title{Hierarchical Structure and Self-gravity in the Maddalena Giant Molecular Cloud}

\correspondingauthor{Yuehui Ma}
\email{mayh@pmo.ac.cn}

\author[0000-0003-2254-1206]{Renjie Shen}
\affiliation{Purple Mountain Observatory and Key Laboratory of Radio Astronomy, Chinese Academy of Sciences, 10 Yuanhua Road, Nanjing 210033, People's Republic of China}
\affiliation{School of Astronomy and Space Science, University of Science and Technology of China, 96 Jinzhai Road, Hefei 230026, People's Republic of China}

\author[0000-0002-8051-5228]{Yuehui Ma}
\affiliation{Purple Mountain Observatory and Key Laboratory of Radio Astronomy, Chinese Academy of Sciences, 10 Yuanhua Road, Nanjing 210033, People's Republic of China}

\author[0000-0003-0746-7968]{Hongchi Wang}
\affiliation{Purple Mountain Observatory and Key Laboratory of Radio Astronomy, Chinese Academy of Sciences, 10 Yuanhua Road, Nanjing 210033, People's Republic of China}
\affiliation{School of Astronomy and Space Science, University of Science and Technology of China, 96 Jinzhai Road, Hefei 230026, People's Republic of China}



\author[0009-0004-2947-4020]{Suziye He}
\affiliation{Purple Mountain Observatory and Key Laboratory of Radio Astronomy, Chinese Academy of Sciences, 10 Yuanhua Road, Nanjing 210033, People's Republic of China}
\affiliation{School of Astronomy and Space Science, University of Science and Technology of China, 96 Jinzhai Road, Hefei 230026, People's Republic of China}

\author[0000-0002-6388-649X]{Miaomiao Zhang}
\affiliation{Purple Mountain Observatory and Key Laboratory of Radio Astronomy, Chinese Academy of Sciences, 10 Yuanhua Road, Nanjing 210033, People's Republic of China}



\begin{abstract}

In this work, we present the data from the Milky Way Imaging Scroll Painting (MWISP) project for the Maddalena giant molecular cloud (GMC). We decompose the $^{13}$CO emission datacube of the observed region into hierarchical substructures using a modified Dendrogram algorithm. We investigate the statistical properties of these substructures and examine the role that self-gravity plays on various spatial scales. The statistics of the mass (M), radius (R), velocity dispersion ($\sigma_v$), virial parameter ($\alpha_{vir}$), and sonic Mach number of the substructures are presented. The radius and mass distributions and the $\sigma_v$-R scaling relationship of the substructures resemble those reported in previous studies that use non-hierarchical algorithms to identify the entities. We find that for the hierarchical substructures $\alpha_{vir}$ decreases as the radius or mass of the substructures increases. The majority of the substructures in the quiescent region of Maddalena GMC are not gravitationally bound ($\alpha_{vir}>2$), while most of the substructures in the star-forming regions are gravitationally bound ($\alpha_{vir}<2$). Furthermore, we find that self-gravity plays an important role on scales of 0.8-4 pc in the IRAS 06453 star-forming region, while it is not an important factor on scales below 5 pc in the non-star-forming region.


\end{abstract}




\section{Introduction}\label{sec1}
Stars form through the gravitational collapse of molecular cores. However, the star formation rate should be much higher than current observations if molecular clouds undergo free-fall collapse. This discrepancy suggests that other physical processes, such as turbulence and magnetic fields, are acting to prevent the rapid gravitational collapse of molecular clouds. Theoretical models propose that turbulence functions as a regulatory process capable of both supporting molecular clouds against self-gravity on large scales (ranging from sub-parsec to parsec) and creating local overdense regions within clouds \citep{2004RvMP...76..125M,2007ARA&A..45..565M}. It is known that self-gravity dominates the evolution of molecular cores smaller than 0.1 pc \citep{2007prpl.conf...17D}, while turbulent fragmentation can produce the observed statistical properties of molecular clumps \citep{2007ApJ...657..870V}. \citet{2009Natur.457...63G} analyzed the structure of the L1448 molecular cloud on different spatial scales and showed that self-gravity plays an important role up to 1 pc. Indeed, observations have shown that molecular clouds and filaments are undergoing global multi-scale collapse on scales less than $\sim$ 1 pc \citep{2009ApJ...706.1036G,2013A&A...555A.112P,2015A&A...581A.119B,2015ApJ...804...37L,2019MNRAS.490.3061V}. However, the importance of self-gravity on larger scales (above $\sim$ 1 pc) is still unclear. For example, the observed Larson's velocity dispersion-size relation spanning from sub-parsec to parsec scales can be explained by compressible turbulence without gravity, either in theoretical studies or numerical simulations \citep{2011arXiv1111.2827K, 2011IAUS..270..179K}, while the gravitational energy spectra of the Orion A and the Orion B clouds derived from observational data show that gravity is dominant for the evolution of these clouds from cloud scale down to $\sim0.1$ pc \citep{2017MNRAS.464.4096L}. 

The internal hierarchical structure of molecular clouds on various scales shed light on the dominant physical process in the evolution of molecular clouds and star formation. Previously, statistics of the physical properties of molecular clouds on multi-scale were often obtained by examining the samples of molecular clouds, clumps, or cores using segmentation algorithms such as Gaussclumps \citep{1990ApJ...356..513S}, Clumpfind \citep{1994ApJ...428..693W}, and Fellwalker \citep{2015A&C....10...22B}. However, these traditional algorithms do not treat molecular clouds as hierarchical structures, which should be natural outcomes of the interplay between turbulence and gravity. Dendrogram \citep{2008ApJ...679.1338R} is a structure-identification algorithm which can reveal the underlying hierarchy in 2-dimensional (2D) or 3-dimensional (3D) datasets. It is not only an effective core/clump identification algorithm \citep{2020RAA....20...31L} but also a practical tool for investigating turbulence \citep{2013ApJ...770..141B}. 

The G216$-$2.5, also known as Maddalena's cloud, was initially discovered by \citet{1985ApJ...294..231M} in a CO ($J=1-0$) survey of the Orion-Monoceros region. This giant molecular cloud has physical dimensions of 250$\times$100 pc$^2$ and a total mass between 1$-6\times 10^5M_{\odot}$ \citep{1985ApJ...294..231M}, situated at a distance of 2.41 kpc \citep{2019ApJ...885...19Y}. The Maddalena GMC was once considered a cold and quiescent molecular cloud with a low kinetic temperature of 10 K and barely any star-formation activity \citep{1985ApJ...294..231M}. \citet{1994ApJ...432..167L}  found the presence of shell-like structures in the Maddalena GMC and suggested that Maddalena GMC is a remnant cloud from a past episode of massive star formation. However, \citet{1996ApJ...472..275L} revealed two small groups of young stars in satellite clouds on the northern edge of the Maddalena GMC and \citet{2009AJ....137.4072M} identified a population of 41 young stars with disks and 33 protostars in the center of the cloud, indicating that the central part of the Maddalena GMC is actively forming stars. \citet{1998ApJ...494..657W} investigated the high-density gas contents in the central part of the Maddalena GMC through observations of multitransition CO and CS(2-1) lines and found a deficiency of dense gas in the Maddalena central region relative to the Rosette GMC. The role of turbulent fragmentation in regulating the star formation efficiency in the Maddalena GMC has been studied by \cite{2006ApJ...643..956H}, using the $^{12}$CO and $^{13}$CO ($J=1-0$) data from the 14 m telescope of the Five College Radio Astronomy Observatory. \cite{2019ApJ...885...19Y} present $^{12}$CO and $^{13}$CO ($J=1-0$) maps toward this region from the Milky Way Imaging Scroll Painting (MWISP) survey \citep{2019ApJS..240....9S}. However, \cite{2019ApJ...885...19Y} are focused on the distance estimations for molecular clouds in the third quadrant of the Milky Way mid-plane and did not perform detailed investigations on the structures and properties of the Maddalena GMC. The neighboring molecular clouds associated with the S287 \ion{H}{2} region, which are at a distance of 2.19 kpc, have been investigated in detail by \cite{2016A&A...588A.104G}. As estimated by \cite{2016A&A...588A.104G}, the S287 massive star forming region has a mean column density of about 4$\times10^{21}$ cm$^{-2}$, which is much higher than the mean column density of the Maddalena GMC derived from Herschel observations (7.8$\times10^{20}$ cm$^{-2}$, \citealp{2015ApJ...803...38I}).

In this work, we present the spatial distribution of physical properties of the Maddalena GMC using the data from the MWISP survey. We decompose the $^{13}$CO data into hierarchical sub-structures using the Dendrogram algorithm, then analyze the statistical properties of these sub-structures and explore the significance of self-gravity at various spatial scales within the Maddalena GMC. In Section \ref{sec2}, we briefly introduce the observations, data reduction processes, and the Dendrogram algorithm. The main results are presented in Section \ref{sec3}. In Section \ref{sec4}, we discuss our results and we make conclusions in Section \ref{sec5}.

\section{Observations and Method}\label{sec2}
\subsection{Observations}\label{sec2.1}

The Maddalena GMC observation is part of the MWISP survey, which observes $^{12}$CO, $^{13}$CO, and C$^{18}$O ($J=1-0$) emission in the northern sky from $l=9^{\circ}.75$ to $l=230^{\circ}.25$, and $b=-5^{\circ}.25$ to $b=5^{\circ}.25$ \citep{2019ApJS..240....9S}. Observations toward the Maddalena region were conducted between September 2013 and May 2015 using the PMO$-$13.7m telescope in Delingha, China, which is equipped with a nine-beam Superconducting Spectroscopic Array Receiver (SSAR) working in the 85$-$115 GHz frequency band \citep{2012ITTST...2..593S}. The half-power beamwidth (HPBW) of the telescope is approximately $52''$ and $50''$ at 110 GHz and 115 GHz, respectively. 

The observations were conducted in the position switch on-the-fly (OTF) mode. The Maddalena GMC is located in the northern sky from l=213$.\!\!^{\circ}$5 to l=219$.\!\!^{\circ}$5, and b=$-$4$.\!\!^{\circ}$5 to b=$-$1$^{\circ}$. The sky area is divided into $30'\times30'$ cells. For each cell, the scannings were made along the directions of Galactic longitude and Galactic latitude, respectively. The antenna temperature is calibrated according to $T_{mb} = T^{\ast}_{A} / \eta_{mb}$ during the data reduction processes. The final data are regrided into spatial pixels of sizes of $30\arcsec \times 30 \arcsec$. The final root mean square (RMS) noise level is $\sim$0.5 K per 0.16 km s$^{-1}$ at the $^{12}$CO ($J=1-0$) line wavelength and $\sim$0.3 K per 0.17 km s$^{-1}$ at the $^{13}$CO ($J=1-0$) and C$^{18}$O ($J=1-0$) line wavelength.

\subsection{Method}\label{sec2.2}



The Dendrogram algorithm \footnote{\url{https://github.com/dendrograms/astrodendro}} \citep{2008ApJ...679.1338R} decomposes input data into hierarchical structures using three user defined parameters, $min\_value$, $min\_ delta$, and $min\_ npix$. The algorithm begins at the maximum within the input data and proceeds by gradually decreasing the intensity level and searching for other local maxima.  When two substructures merge at a certain level, the algorithm verifies if the number of voxels, $n_{voxel}$, within each substructure exceeds $min\_npix$. Additionally, it checks if the intensity difference between the local maximum and the merging level (denoted as $\delta$) exceeds $min\_delta$. The substructures that satisfy $n_{voxel} > min\_npix$ and $\delta>min\_delta$ are considered independent structures, and are referred to as ``leaves'', otherwise, they are merged into a larger structure, which is referred to as ``branch''. The algorithm stops searching for new local maxima and associated substructures once the current intensity of the isosurface reaches $min\_ value$. The structures directly originating from $min\_value$, either ``branches'' or ``leaves'', are termed ``trunks'' and are marked as structures at ``level 0''. Subsequently, the direct offspring of these trunks are regarded as structures at ``level 1'', and this pattern continues. Conversely, we can define the stages for the structures starting from the ``leaves''. In this scenario, ``leaves'' are considered as structures at ``stage 0'', while their immediate parent ``branches'' are at ``stage 1''. Parental ``branches'' of ``stage 1'' substructures belong to ``stage 2'', and so forth. In short, ``leaves'' merge to form ``branches'', and ``branches'' merge to constitute a ``tree''.

To characterize the identified dendrogram tree, we define the peak brightness temperatures as the ``height'' of the associated substructure. The difference in brightness temperatures between local maximum and merging node is designated as the ``length'' of the leaf. The difference between the merging nodes is defined as the ``length'' of the branch.

The original ``Astrodendro'' algorithm forces the formation of a binary tree, i.e., one branch has only two children, which is not necessarily the hierarchical structure of molecular cloud. For example, arbitrary numbers of clumps or cores could merge at a specific intensity level to form a cloud. Therefore, modifications to the original algorithm have been made to allow arbitrary numbers of children to merge into one branch. A new constraint parameter, $branch$\_$delta$, is introduced to merge the branches whose lengths are smaller than the noise level of the data. This modification ensures that the branches are physically meaningful structures. Details on the algorithm modification are given in a forthcoming paper by S. He et al. 2024 (in prep).


In this work, we identify hierarchical structures in the $^{13}$CO data and set the parameters $min\_value$, $min\_ delta$, $branch\_delta$ and $min\_npix$ to be 3$\sigma_{13,\rm RMS}$, 2$\sigma_{13,\rm RMS}$, 1$\sigma_{13,\rm RMS}$ and 27 voxel, respectively.

\section{Results}\label{sec3}

\subsection{Overall Structure and Properties of the Maddalena GMC} \label{sec3.1}
\begin{figure}
  \gridline{\fig{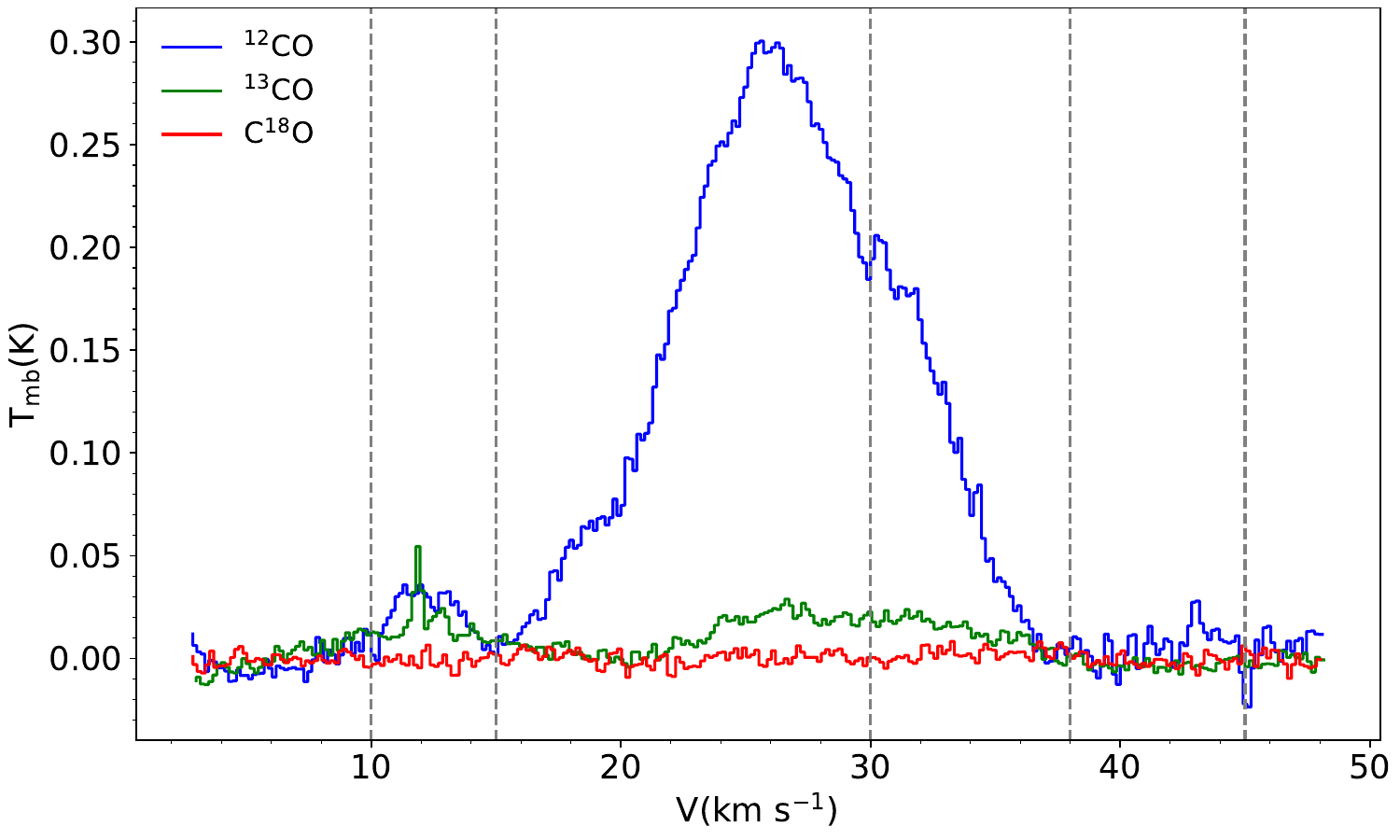}{0.48\textwidth}{(a)}
            }
  \gridline{\fig{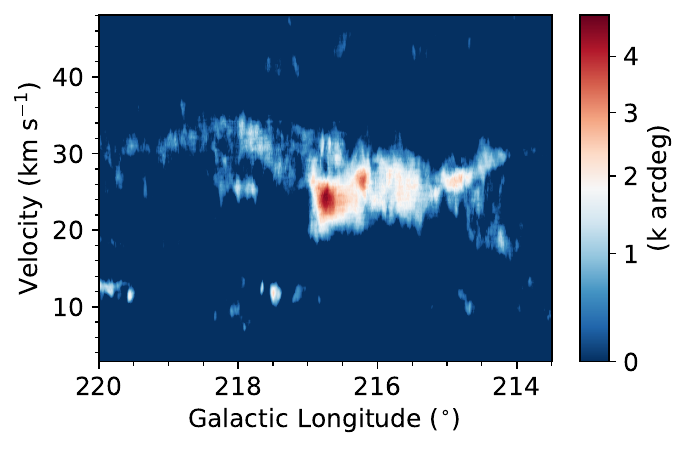}{0.5\textwidth}{(b)}
            }
  \caption{(a) Average spectra of the Maddalena region with an area of $6^{\circ} \times 3.\!\!^{\circ}5$, where the blue, green, and red represent the emission of ${ }^{12} \mathrm{CO},{ }^{13} \mathrm{CO}$, and $\mathrm{C}^{18} \mathrm{O}$, respectively. The velocity range of the emission lines in the ${ }^{12} \mathrm{CO}$ spectrum can be roughly divided into four parts (10 to 15, 15 to 30, 30 to 38, and 38 to 45 km $\mathrm{s}^{-1}$), represented by vertical dashed lines. (b) Galactic longitude $-$ velocity ($l-v$) diagram of $^{12}$CO emission integrated along the Galactic latitude from b=$-$4$.\!\!^{\circ}$5 to b=$-$1$^{\circ}$.}
  \label{fig2}
\end{figure}
Figure \ref{fig2}(a) shows the average spectra of the $l$ = 213$.\!\!^{\circ}5$ to 219$.\!\!^{\circ}5$ and $b$ = $-4.\!\!^{\circ}5$ to $-1^{\circ}$ region which covers nearly the whole Maddalena GMC. The observed area also covers the two satellite clouds of Maddalena GMC, associated with the IRAS 06453 and IRAS 06522 sources identified by \cite{1996ApJ...472..275L}. The two satellite molecular clouds are different from the Maddalena GMC in terms of more active star formation and the existence of young stellar groups \citep{1996ApJ...472..275L, 2009AJ....137.4072M}. In Figure \ref{fig2}(a), the blue, green, and red represent the emission of ${ }^{12} \mathrm{CO},{ }^{13} \mathrm{CO}$, and $\mathrm{C}^{18} \mathrm{O}$, respectively. Among these spectral lines, ${ }^{12} \mathrm{CO}$ radiation has the highest brightness temperature, while $\mathrm{C}^{18} \mathrm{O}$ has the lowest. We found that at very few positions in the observed region, the $\mathrm{C}^{18} \mathrm{O}$ spectra meet the criterion of containing at least three contiguous velocity channels with brightness temperature higher than $2 \sigma$. The velocity range of the average ${ }^{12} \mathrm{CO}$ spectrum can be roughly divided into four parts, i.e., 10 to 15, 15 to 30, 30 to 38, and 38 to 45 km $\mathrm{s}^{-1}$, respectively, represented by vertical dashed lines in Figure \ref{fig2}(a). Figure \ref{fig2}(b) shows the Galactic longitude-velocity distribution of the $^{12}$CO emission. In Figure \ref{fig2}(b), the Maddalena GMC is identified as a velocity coherent structure within the velocity range from 15 to 38 km $\mathrm{s}^{-1}$. There exist some fragmented structures at velocities of around 10-12 km s$^{-1}$ and beyond 40 km s$^{-1}$. These structures are not connected to the Maddalena GMC either in velocity or in sky projection. Therefore, in this work, we present maps derived within the velocity range of [15, 38] km s$^{-1}$ and analyze the tree structure of the Maddalena GMC within this range. A more detailed description of Figure \ref{fig2}(b) will be provided in Section \ref{sec3.1.2} where we discuss the velocity structure of the Maddalena GMC.

\subsubsection{Spatial Distribution of the Maddalena GMC} \label{sec3.1.1}

\begin{figure*}
  \centering
    \begin{minipage}[t]{0.8\linewidth}
    \centering
    \subfigure[]{
     \raisebox{0.05\height}{\includegraphics[width=\textwidth]{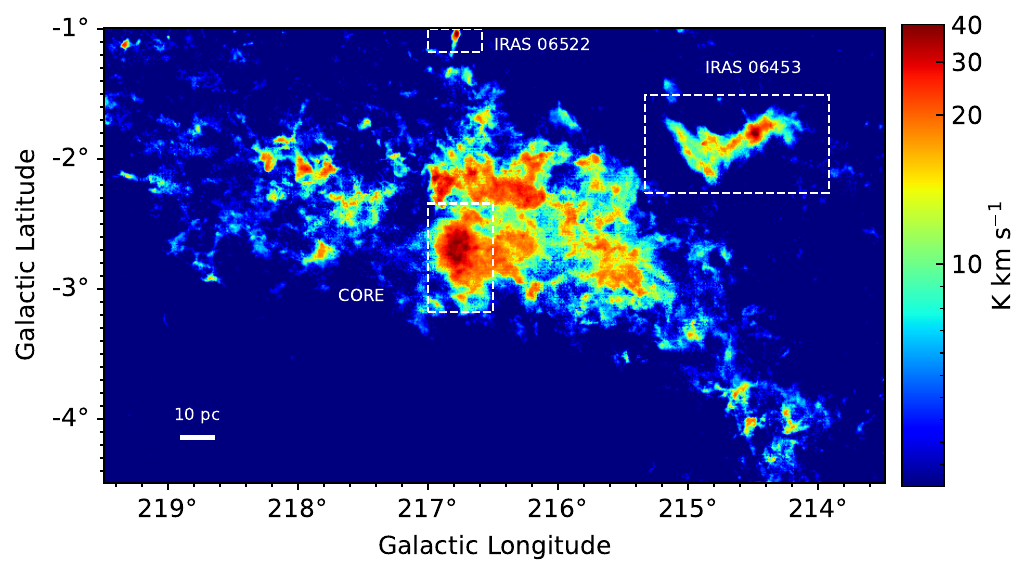}}
    }
   \end{minipage}
   \begin{minipage}[t]{0.8\linewidth}
    \centering 
    \subfigure[]{
     \raisebox{0.05\height}{\includegraphics[width=\textwidth]{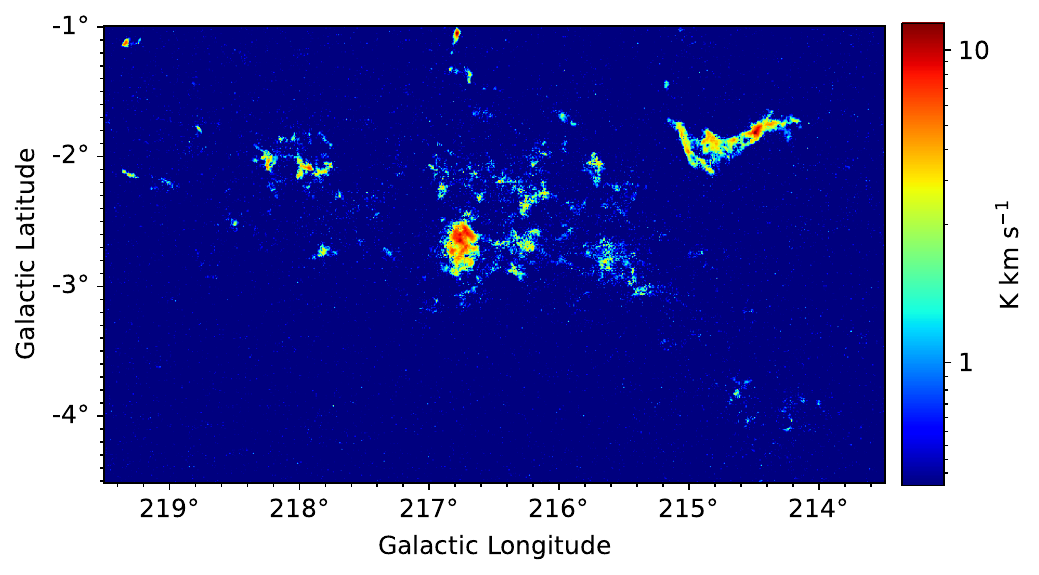}}
    }
   \end{minipage}
  \caption{Integrated intensity maps of (a) $^{12}$CO and (b) $^{13}$CO emission. The velocity integration range is from 15 to 38 km $\mathrm{s}^{-1}$. The three regions discussed in Section \ref{sec3.1.1}, i.e., Maddalena core, IRAS 06453, and IRAS 06522, are marked with white dashed rectangles in panel (a).}
  \label{fig3}
\end{figure*}

The integrated intensity maps of ${ }^{12} \mathrm{CO}$ and $^{13}$CO $(J=1-0)$ emissions in the region are shown in Figure \ref{fig3}. Panel (a) displays the integrated intensity map for $^{12}$CO data, which have a maximum value of 56.1 K km $\mathrm{s}^{-1}$. This peak integrated intensity is observed within the core region of the Maddalena molecular cloud. In the IRAS 06453 molecular cloud, the integrated intensity values also reach a high level. The IRAS 06522 cloud shows as a bright compact core. Compared to the typically optically thick ${ }^{12} \mathrm{CO}$ emission, ${ }^{13} \mathrm{CO}$ emission is usually optically thin and is commonly used to trace denser regions of molecular clouds. The maximum integrated intensity value of the $^{13}$CO emission in the region is 12.8 K km $\mathrm{s}^{-1}$. The spatial distribution of ${ }^{13} \mathrm{CO}$ integrated intensities shown in Figure \ref{fig3} (b) is generally consistent with that of ${ }^{12} \mathrm{CO}$, but with lower values. In comparison to ${ }^{12} \mathrm{CO}$ and ${ }^{13} \mathrm{CO}$, $\mathrm{C}^{18}\mathrm{O}$ traces molecular clouds with higher column densities. In our data, the $\mathrm{C}^{18}\mathrm{O}$ emission is very weak, barely tracing any contiguous structures but only some scattered points with intensities of around 0.6 K km $\mathrm{s}^{-1}$. We attribute the weak C$^{18}$O emission in the region to the relatively low column density of the Maddalena cloud, which is presented in Section \ref{sec3.1.3}. For comparison, \cite{2009Natur.457...63G} have performed dendrogram analysis for the L1448 region which possesses much higher column density and shows significant C$^{18}$O emission \citep{2005A&A...440..151H}.
  


\begin{figure*}
  \centering
   \begin{minipage}[t]{0.7\linewidth}
    \centering
    \subfigure[]{
     \raisebox{0.05\height}{\includegraphics[width=\textwidth]{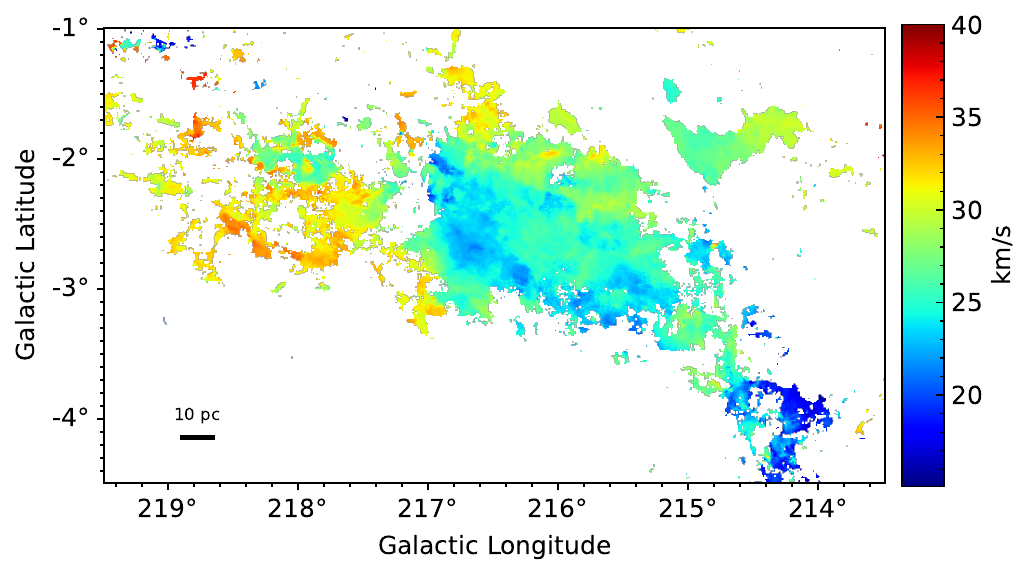}}
    }
   \end{minipage}
   \begin{minipage}[t]{0.7\linewidth}
    \centering 
    \subfigure[]{
     \raisebox{0.05\height}{\includegraphics[width=\textwidth]{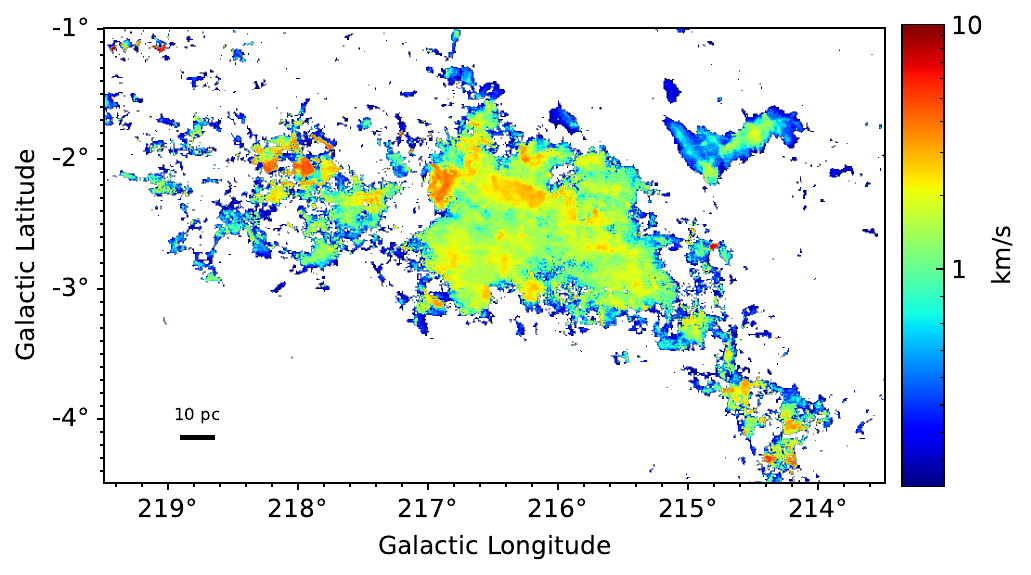}}
    }
   \end{minipage}
  \caption{(a) Centroid velocity map of $^{12}$CO emission. (b) Velocity dispersion map of $^{12}$CO emission. }
  \label{fig4}
\end{figure*}

\subsubsection{Velocity Structure of the Maddalena GMC} \label{sec3.1.2}
The velocity distribution of $^{12}$CO emission of the Maddalena GMC within the velocity range from 15 to 38 km $\mathrm{s}^{-1}$ is shown in Figure \ref{fig4} (a). From Figure \ref{fig4} (a), we can see that the Maddalena GMC exhibits a large-scale velocity gradient in the direction from northeast to southwest within the Galactic coordinate system, with redshifted velocities observed in the northeastern part and blueshifted velocities in the southwestern part of the GMC. At a distance of 2.41 kpc, the gradient is about $\sim$ 0.08 km s$^{-1}$ pc$^{-1}$. The $^{13}$CO velocity distribution within the GMC closely resembles that of the $^{12}$CO velocity. Considering the narrower spatial distribution of $^{13}$CO emission than $^{12}$CO emission, we do not present the centroid velocity map of $^{13}$CO in Figure \ref{fig4}. The velocity dispersion of the Maddalena molecular cloud is shown in Figure \ref{fig4} (b). We can see that the maximum velocity dispersion from the $^{12}$CO spectra of the Maddalena GMC is $\sim$6 km $\mathrm{s}^{-1}$, showing in the central part of the GMC. The velocity dispersions at the central part of the GMC are, overall, higher than that at the outer layer of the GMC. The velocity dispersions of a few pixels at the northeastern corner of the map can reach up to $\sim$10 km s$^{-1}$. However, these are caused by multiple velocity components within the velocity range of 15-38 km s$^{-1}$. The velocity gradient seen in Figure \ref{fig4} (a) is also evident in Figure \ref{fig2}(b). In Figure \ref{fig2}(b), there are two emission peaks which are concentrated around velocity 25 km $\mathrm{s}^{-1}$. The left emission peak corresponds to the central core region of the Maddalena GMC, which once was considered quiescent until \citet{2009AJ....137.4072M} found vigorous star formation within it. The core region not only shows higher emission intensity but also exhibits broader velocity dispersion than other regions of the GMC.

\begin{figure*}
  \centering
  \begin{minipage}[t]{0.75\linewidth}
   \centering
   \subfigure[]{
    \raisebox{0.05\height}{\includegraphics[width=\textwidth]{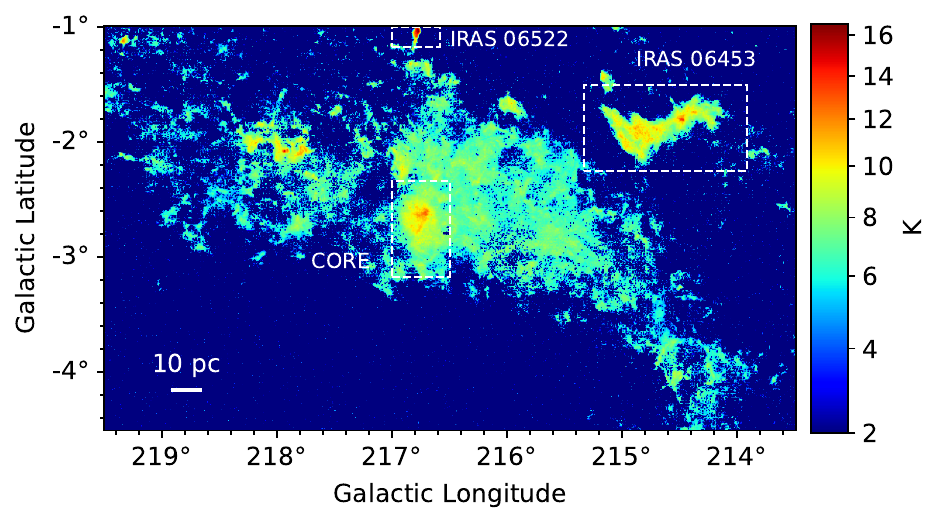}}
   }
  \end{minipage}
  \begin{minipage}[t]{0.75\linewidth}
   \centering 
   \subfigure[]{
    \raisebox{0.05\height}{\includegraphics[width=\textwidth]{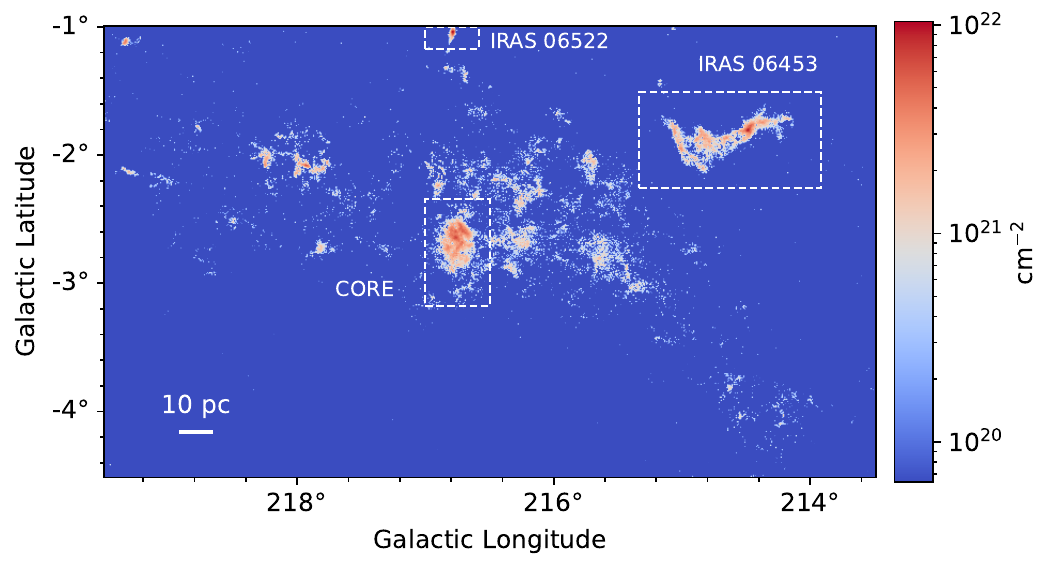}}
   }
  \end{minipage}
  \caption{(a) Excitation temperature map derived from the $^{12}$CO data. (b) H$_2$ column density map calculated from the $^{13}$CO data assuming the LTE condition.} 
  \label{fig5}
 \end{figure*}

 \begin{figure*}
  \centering
  \begin{minipage}[t]{0.8\linewidth}
   \centering
   \subfigure[]{
    \raisebox{0.05\height}{\includegraphics[width=\textwidth]{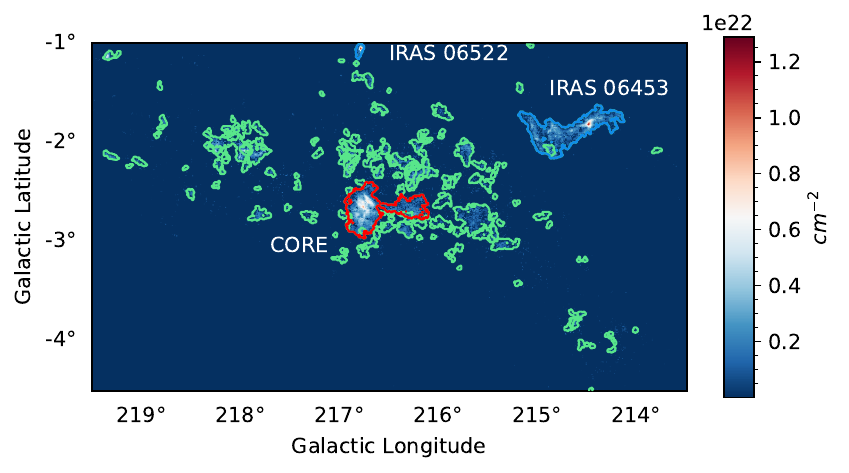}}
   }
    \end{minipage}
  \begin{minipage}[t]{0.7\linewidth}
   \centering 
   \subfigure[]{
    \raisebox{0.05\height}{\includegraphics[width=\textwidth]{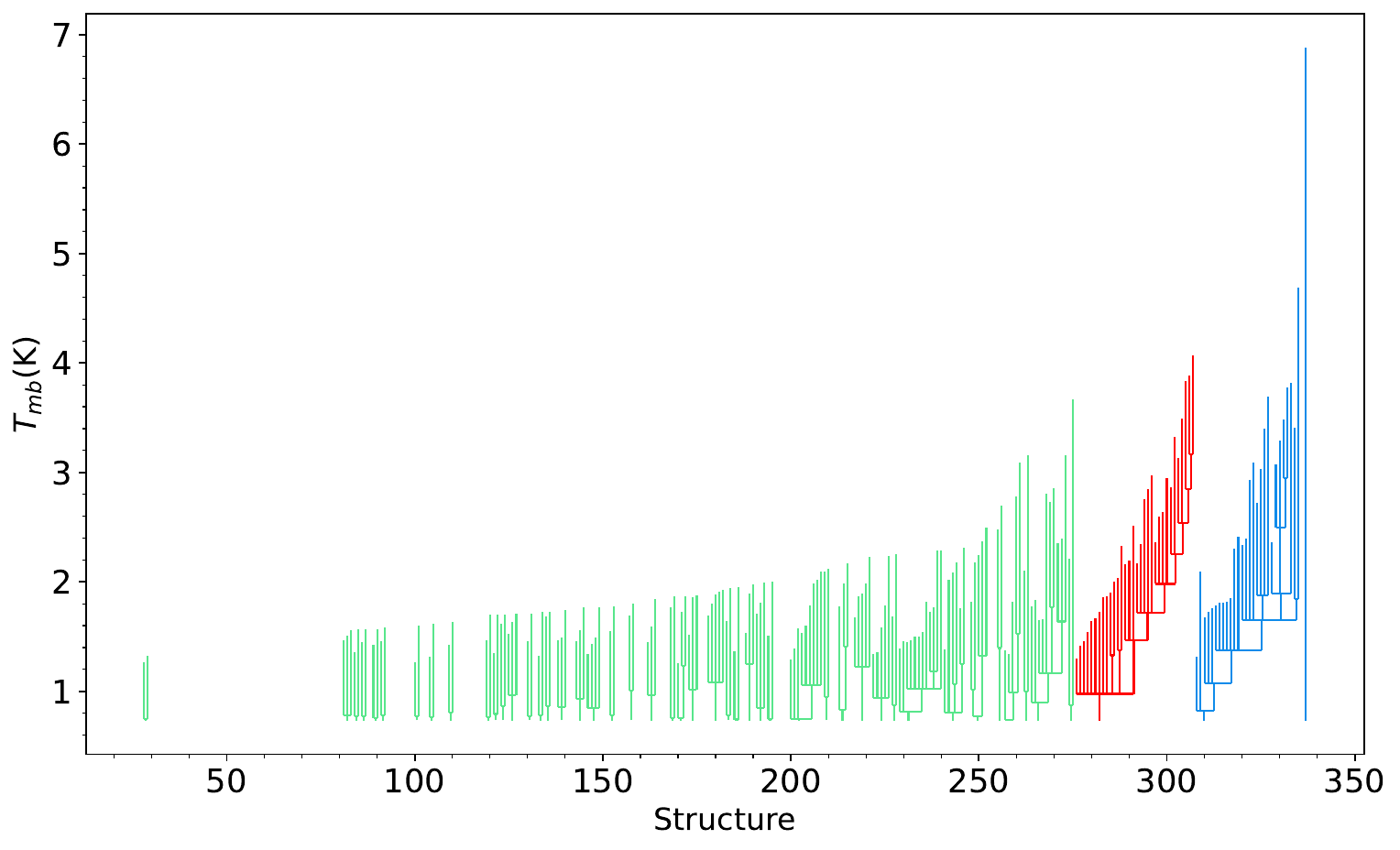}}
   }
  \end{minipage}
     \caption{(a) Contours of structures identified in the 3D $^{13}$CO datacube overlaid on the H$_2$ column density map. (b) Tree diagram of the structures identified by the Dendrogram algorithm. Two satellite star-forming clouds IRAS 06522 and IRAS 06453 \citep{2009AJ....137.4072M} are highlighted in blue in both panels. The central core of the Maddalena GMC is highlighted in red. Other structures are highlighted in green. }
     \label{fig6}
 \end{figure*}

 \subsubsection{Excitation Temperature and Column Density Distribution} \label{sec3.1.3}
 Figure \ref{fig5} shows the excitation temperature ($T_{\rm ex}$) and column density ($N_{H_2}$) maps derived from the $^{12}$CO and $^{13}$CO data. The excitation temperature and the column density of the $^{13}$CO molecule are derived assuming that the $^{12}$CO and $^{13}$CO molecules are under the local equilibrium state (LTE) and the $^{12}$CO emission is optically thick. Once the $^{13}$CO column density is calculated, the H$_2$ column density can be obtained using the abundance ratios [$^{12}$C]/[$^{13}$C] = 6.21d$_{GC}$+18.71 and [H$_2$]/[$^{12}$CO] = 1.1$\times$10$^4$ \citep{1982ApJ...262..590F,2005ApJ...634.1126M,2016A&A...588A.104G}. Formulas used to derive the excitation temperature and the column density of the H$_2$ molecule are given in Section \ref{sec:maths} in the Appendix. Panel (a) of Figure \ref{fig5} shows the spatial distribution of $T_{\rm ex}$. The Maddalena GMC has a peak excitation temperature of 16.5 K, and for most of the region the excitation temperature is below 10 K. The $^{13}$CO emission is moderately optically thin ($\tau_{13}<1$) in 99.8$\%$ of the observed regions, with $\tau_{13}$ lying in the range from 0.12 to 1.72 and having a median value of 0.34. In Figure \ref{fig5}, the entire GMC exhibits quite low H$_2$ column densities ($6.4\times10^{19}- 1.04\times10^{22}$ cm$^{-2}$), with a median value of $6.6\times10^{20}$ cm$^{-2}$. High column densities, $3\times10^{21}-1.04\times10^{22}$ cm$^{-2}$, exist in the central core region of the main body of the Maddalena GMC and the two satellite star-forming clouds IRAS 06522 and IRAS 06453. Other regions have relatively low column densities, $6.4\times10^{19}- 1\times10^{21}$ cm$^{-2}$, and barely show evidence of star formation.  
 
 \begin{figure*}
  \centering
  \begin{minipage}[t]{0.49\linewidth}
   \centering
   \subfigure[]{
   \raisebox{0.05\height}{\includegraphics[width= \linewidth]{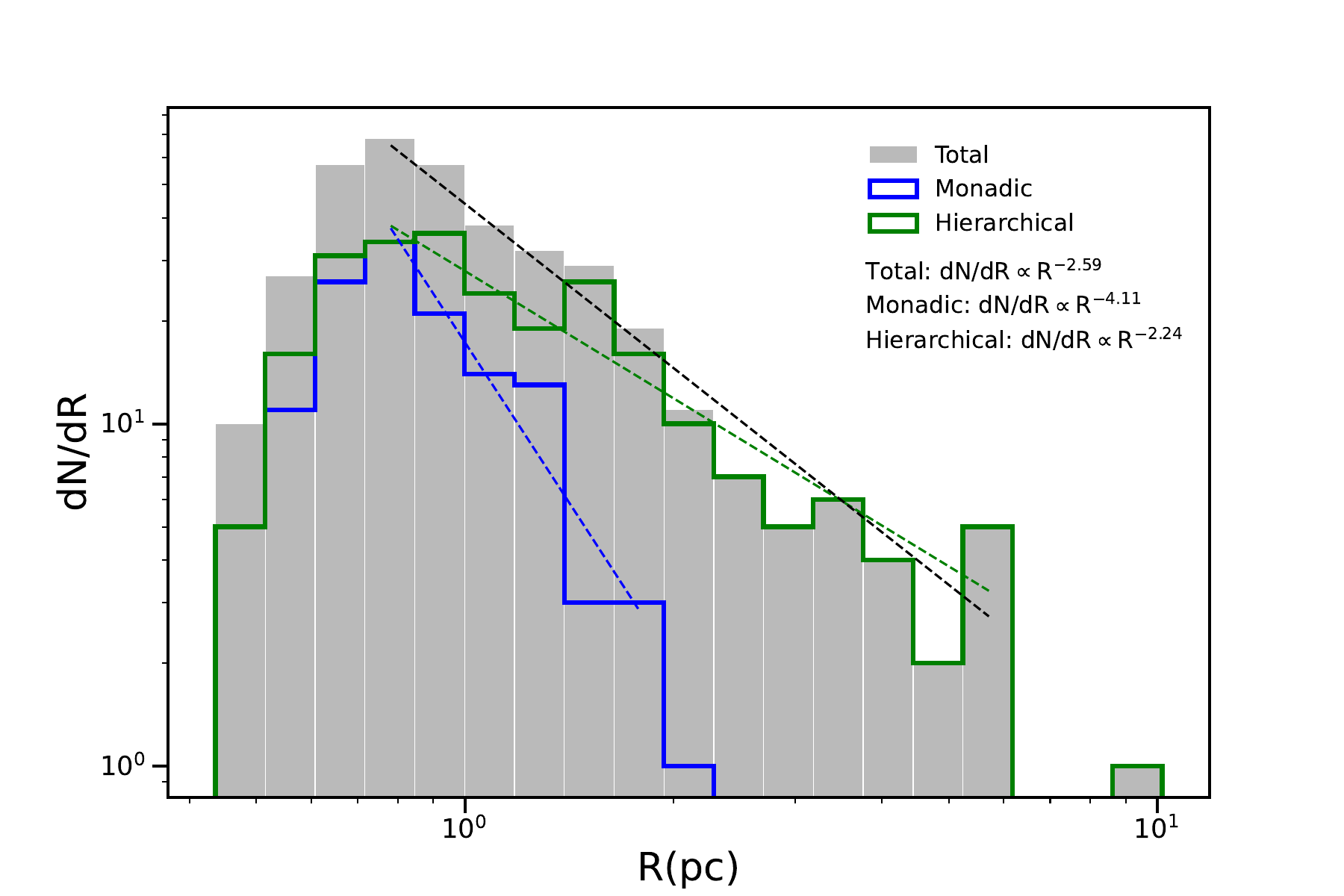}}}
 
  \end{minipage}
  \begin{minipage}[t]{0.49\linewidth}
   \centering 
   \subfigure[]{
   
   \raisebox{0.05\height}{\includegraphics[width= \linewidth]{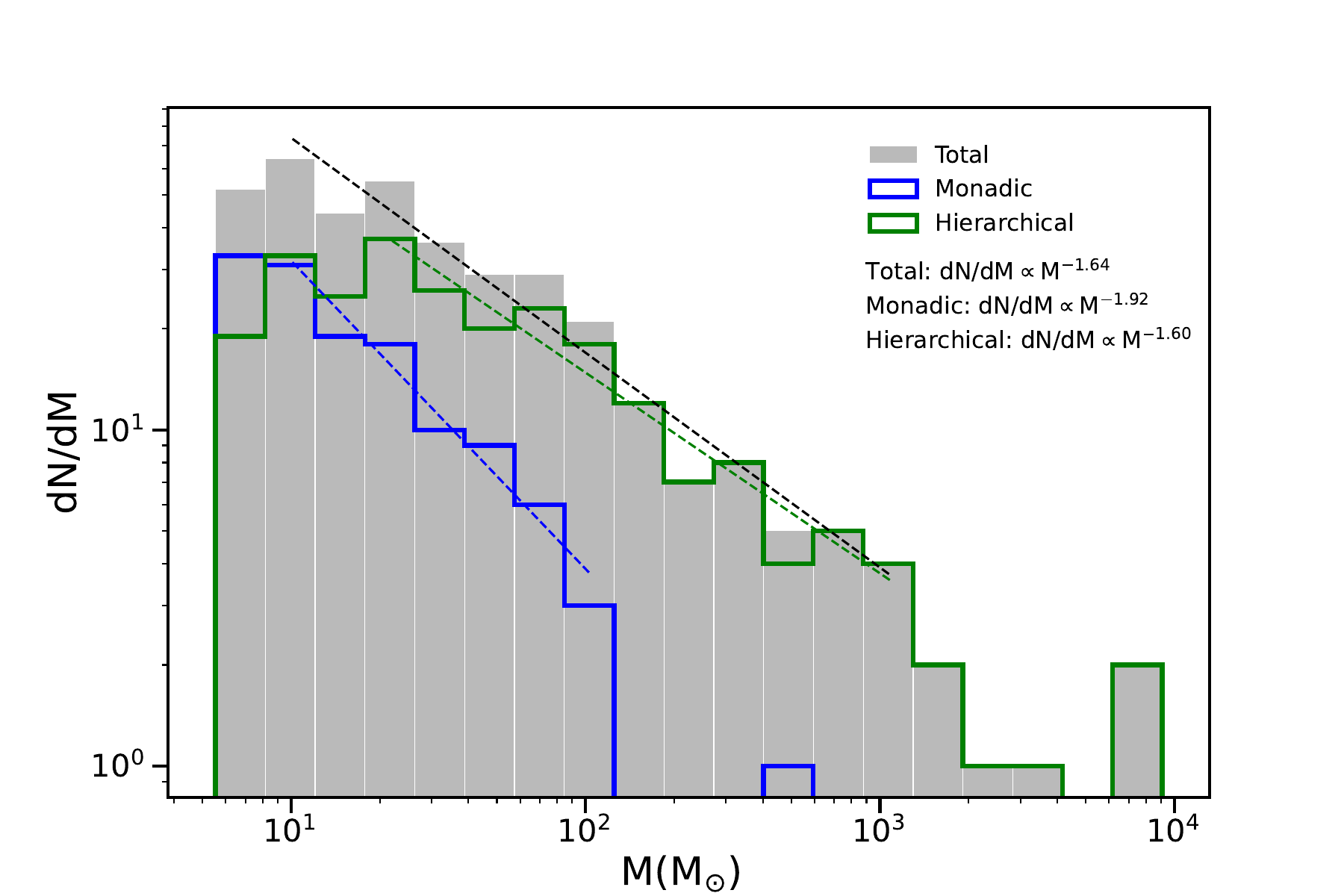}}}
  \end{minipage}
   \begin{minipage}[t]{0.49\linewidth}
   \centering 
   \subfigure[]{
   
   \raisebox{0.05\height}{\includegraphics[width= \linewidth]{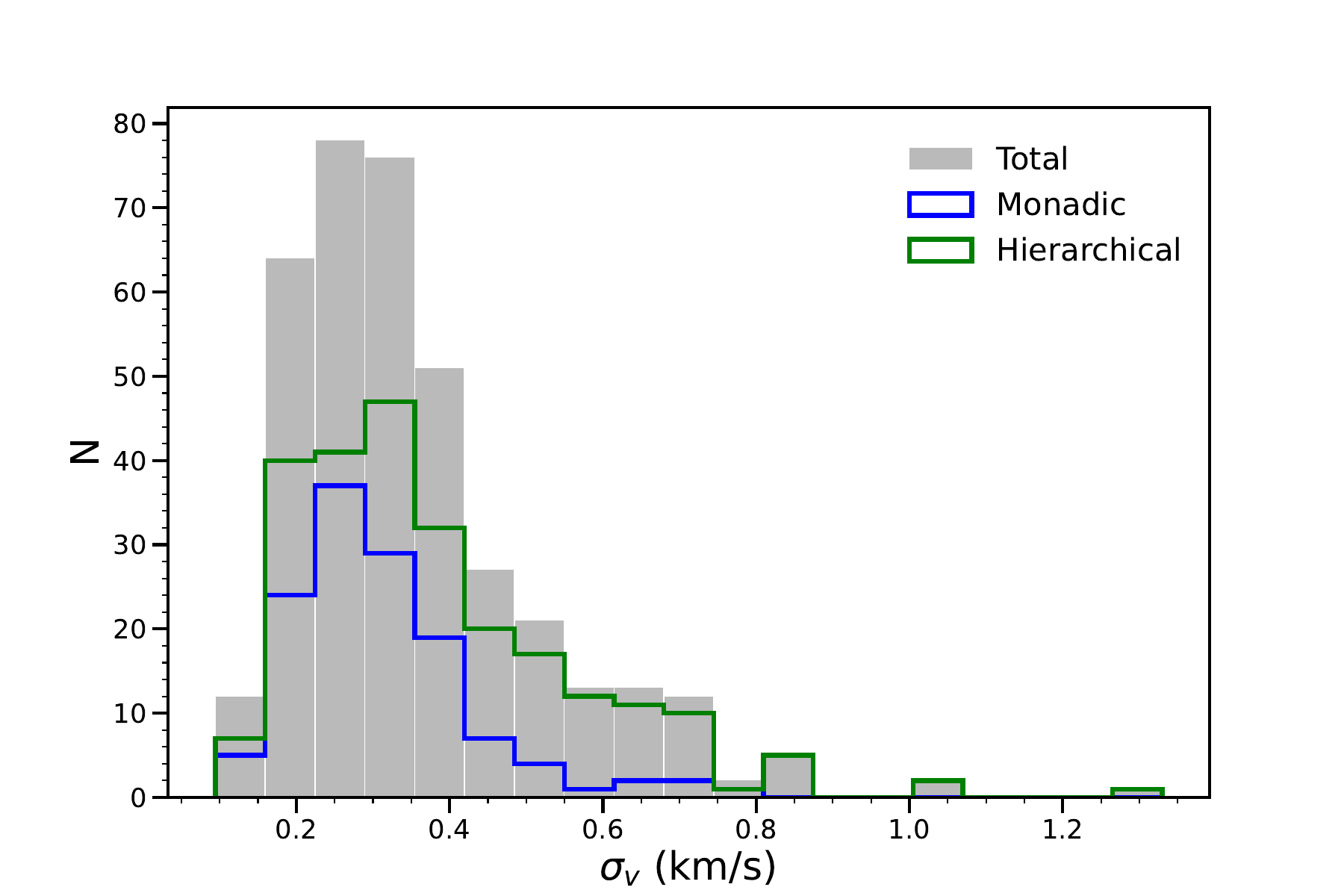}}}
  \end{minipage}
   \begin{minipage}[t]{0.49\linewidth}
   \centering 
   \subfigure[]{
  
   \raisebox{0.05\height}{\includegraphics[width= \linewidth]{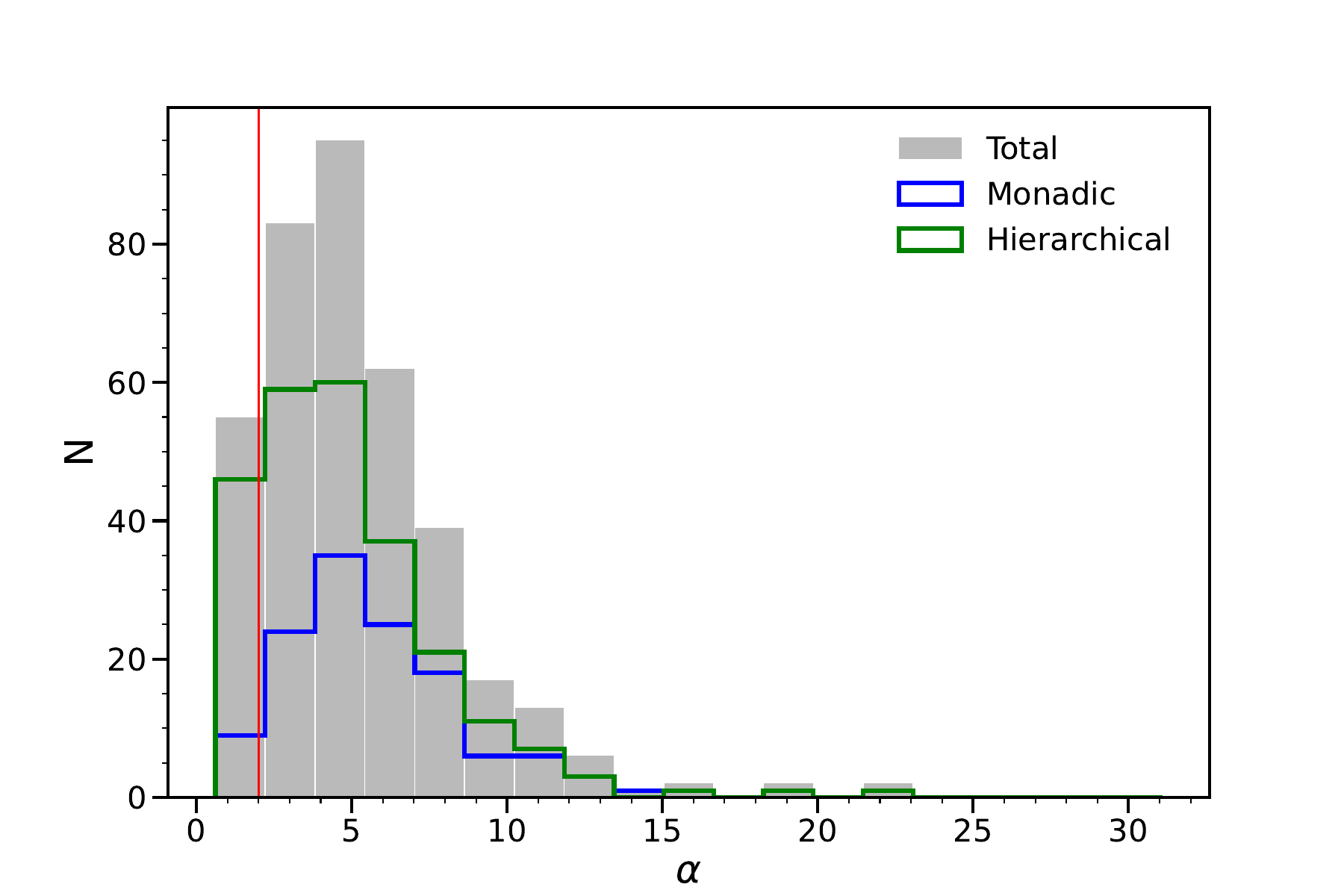}}}
  \end{minipage}
  \caption{Distribution of (a) radius, (b) mass, (c) velocity dispersion, and (d) virial parameter of the hierarchical structures in the Maddalena GMC. The red vertical line in (d) marks the $\alpha$ value of 2.}
  \label{fig7}
 \end{figure*}
 
 \subsection{Hierarchical Structure of the Maddalena GMC} \label{sec3.2}
 
 As introduced in section \ref{sec2.2}, we divided the Maddalena GMC, traced by $^{13}$CO emission, into hierarchical substructures using the modified Dendrogram algorithm.
 The tree diagram of the Maddalena GMC, identified with the parameter set of $min\_value = 3\sigma_{13,\rm RMS}$, $min\_delta = 2\sigma_{13,\rm RMS}$, $branch\_delta = 1\sigma_{13,\rm RMS}$ and $min\_npix = 27$ is illustrated in Figure \ref{fig6}. The chosen parameter of 27 voxels corresponds to a 3$\times$3$\times$3 voxel structure in the PPV space, with dimensions of 90$\arcsec\times$90$\arcsec$ and covering a velocity range of 0.5 km s$^{-1}$. The telescope beam size is 50$\arcsec$. The selected parameter setup ensures that the smallest structures identified have angular sizes at least 1.5 times the beam size after deconvolution, ensuring their physical significance. The $branch\_delta$ parameter is introduced to exclude spurious branching caused by noise in the observational data. A value of 1 $\sigma_{13, RMS}$ for the $branch\_delta$ parameter is sufficient to suppress excessive branching caused by noise while still allowing the algorithm to adequately detect the hierarchical structure of molecular clouds. A larger $branch\_delta$, on the other hand, may cause excessive merging, leading to artificial cutting out of real hierarchical levels. For a concise display, we do not show in Figure \ref{fig6} the trunks in Maddalena GMC that have no substructures. Figure \ref{fig6}(a) shows the boundaries of the identified trees overlaid on the H$_2$ column density map of the GMC. The colors of the contours in panel (a) correspond to the colors in panel (b) of Figure \ref{fig6}. The two satellite star-forming molecular clouds, IRAS 06522 and IRAS 06453, are highlighted in blue in Figure \ref{fig6}. The cloud IRAS 06522 is represented by the right and the highest blue trunk in the diagram which shows no substructure, while IRAS 06453 corresponds to the second highest blue trunk which shows a complex structure. Since these two molecular clouds may have different properties from the Maddalena GMC \citep{1996ApJ...472..275L}, their properties are examined separately from the Maddalena GMC. The central core of the main body is highlighted in red, which corresponds to the third highest trunk in the tree diagram. In total, the Maddalena GMC has 377 substructures. Among the trunks, 132 (73$\%$) are monadic, while 47 (27$\%$) contain at least 2 substructures. For the convenience of the following statistical analyses, we divide the substructures into two groups, i.e., monadic and hierarchical. For additional comparison and discussion, we also divide the Maddalena GMC into two sub-regions, Maddalena-core and Maddalena-quiescent (abbreviated as Maddalena).
 
 \subsection{Statistical Analysis of the Hierarchical Structure of Maddalena GMC} \label{sec3.3}
 \subsubsection{Statistics of Physical Parameters} \label{sec3.3.1}
We calculated the radius, mass, velocity dispersion, and virial parameter for each substructure in the tree, for the details see Section \ref{sec:maths} in the Appendix. In this subsection, we constrain our analysis to the substructures contained in the Maddalena GMC. The histograms of the derived parameters are given in Figure \ref{fig7}. Panel (a) shows the histogram of the radius. At the distance of 2.41 kpc, the angular resolution ($50\arcsec$) of our observations corresponds to a physical resolution of 0.6 pc. Therefore, we only use the fully resolved structures, i.e., those with radii larger than 0.3 pc, to build the histograms in Figure \ref{fig7}.
In panel (a), the distribution of the radius peaks at $\sim$0.7 pc. However, the left side of the distribution is sensitive to the cutoff we used in the structure identification process. Therefore, we fit the radius distribution from its peak to 6 pc. The radius follows a power-law distribution with an exponent of $-$2.59, which is consistent with the results using traditional segmentation algorithms \citep{2001ApJ...551..852H,2015ARA&A..53..583H,2017ApJ...834...57M}. The fitting has a coefficient of determination, $r^2$, of 0.99, indicating a good fit. The substructures in the monadic and hierarchical groups shows clearly different distributions. Specifically, the hierarchical structures have a much shallower power-law distribution, $dN/dR\propto R^{-2.24}$, than the monadic substructures, which have a power-law index of $-$4.11.
 
Figure \ref{fig7}(b) presents the mass distribution ranging from $\sim 5$ to $9\times10^3 M_{\odot}$. We fitted the mass distribution in the range from $10 M_{\odot}$ to $1\times10^3 M_{\odot}$, ensuring that each bin contains at least three counts. It can be seen that the mass of the structures obeys a power-law distribution with a power exponent of $-$1.64, which is similar to the indices found for molecular clouds in the inner galaxy using the traditional segmentation algorithms (-1.6, \citealt{2016ApJ...822...52R}). An index higher than $-$2.0 indicates more masses contained in the high mass end within equally divided logarithmic bins. Similar to the radius distribution, the hierarchical group has shallower mass distribution than that of the monadic group.
 
 \begin{figure}
   \includegraphics[width=\columnwidth]{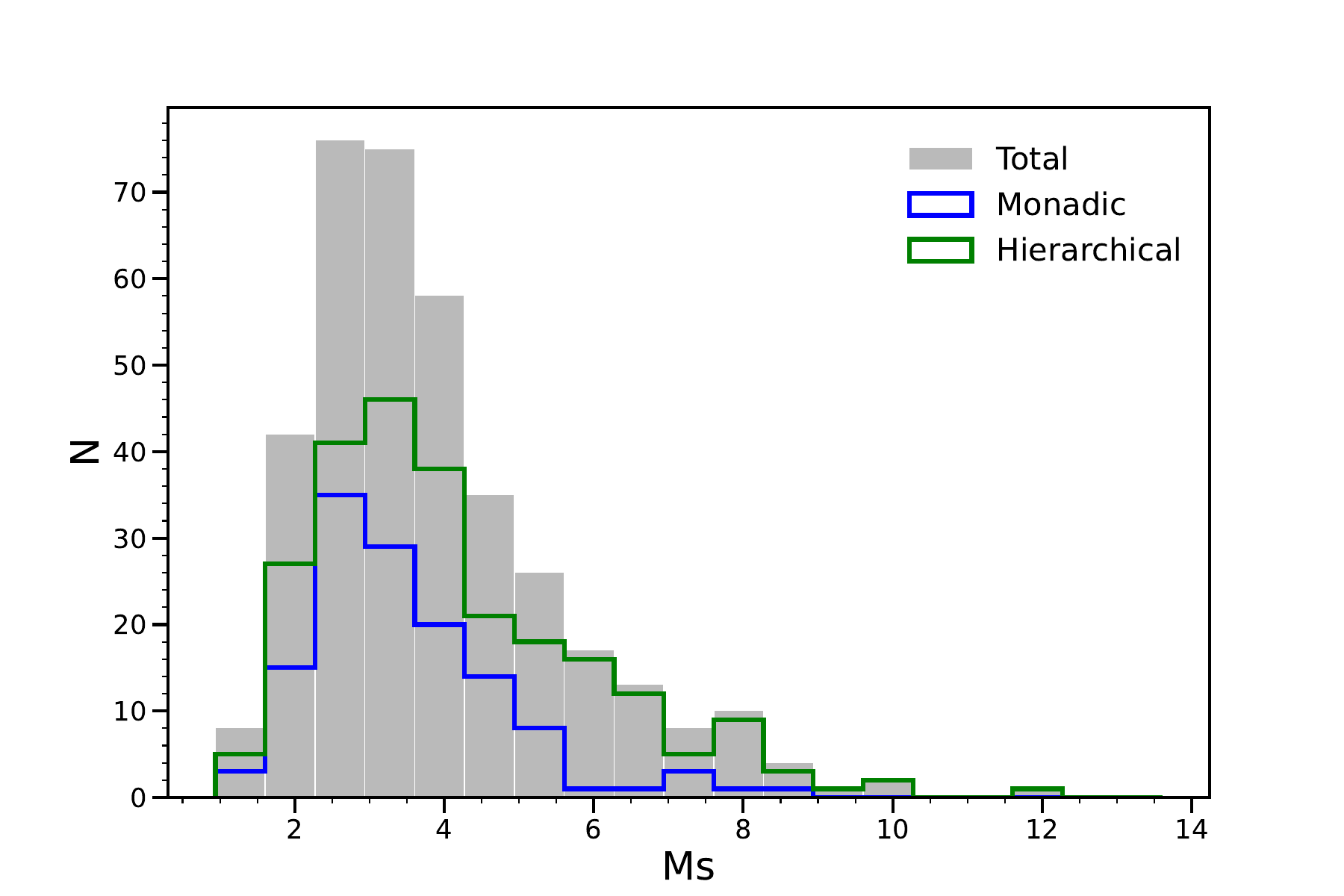}
     \caption{Distribution of Mach numbers.}
     \label{fig8}
 \end{figure}
 
Figure \ref{fig7}(c) shows the velocity dispersion distribution. The observed velocity dispersions range from 0.09 to 1.33 km $\mathrm{s}^{-1}$, with a median value around 0.32 km $\mathrm{s}^{-1}$. Figure \ref{fig7}(d) shows the virial parameter distribution. Structures with virial parameters less than 2 accounts for 11$\%$ of the total, indicating that most of the structures in the Maddalena GMC need external forces to stay in equilibrium. Although the monadic and hierarchical substructures exhibit distinct radius and mass distributions, their distributions of velocity dispersion and virial parameter are quite similar.
 
We also calculated the Mach number of the identified substructures through $\mathcal{M}=\sqrt{3}\sigma_{v_{n o n}} / c_{s}$, where $\sigma_{v_{n o n}}$ is the one-dimensional non-thermal velocity dispersion and $c_{s}$ is the sound speed. The sound speed is derived via $c_{s}=\sqrt{k_{B} T_{k} /\left(\mu_{H} m_{H}\right)}=0.187 \sqrt{T_{k i n} / 10} \mathrm{~km} \mathrm{~s}^{-1}$ \citep{2013ApJ...766L..17S}, where $\mu_{H}=2.37$ is the mean molecular weight per free particle \citep{2008A&A...487..993K}, $m_{H}$ is the mass of the atomic hydrogen, and $k_{B}$ is the Boltzmann constant. The kinetic temperature $T_{kin}$ is approximated to the excitation temperature under the local thermodynamic equilibrium (LTE). The non-thermal velocity dispersion is derived through $\sigma_{v_{non}}=\sqrt{\sigma_{v,{ }^{13} C O}^{2}-c_{s,{ }^{13} C O}^{2}}$, where $c_{s,{ }^{13}CO}=\sqrt{k_{B} T_{k} /\left(\mu_{^{13}CO} m_{H}\right)}$ is the thermal velocity dispersion of $^{13}$CO molecules, and $\mu_{^{13}CO}=29$ is the relative molecular mass of $^{13}$CO molecule. The parameter, $\sigma_{v,^{13}CO}$, is the intensity weighted velocity dispersion which is defined as $\sigma_{v,^{13}CO}=\sqrt{\Sigma\left[T_{i}\left(v_{i}-v_{0}\right)^{2}\right] / \Sigma T_{i}}$, where $v_{0}$ is the intensity-weighted central velocity of the cloud. Figure \ref{fig8} shows the distribution of the derived Mach numbers. The median value of the Mach number is 3.46, 3.25, and 3.64 for the total substructures, the monadic group, and the hierarchical group, respectively, with corresponding standard deviations of 1.86, 1.31, and 2.05. Turbulence in the Maddalena GMC is moderately supersonic for the majority of the structures and does not show significant difference between the monadic and hierarchical substructures.
 
\begin{figure*}
  \centering
  \begin{minipage}[t]{0.32\linewidth}
   \centering
   \includegraphics[width= \linewidth]{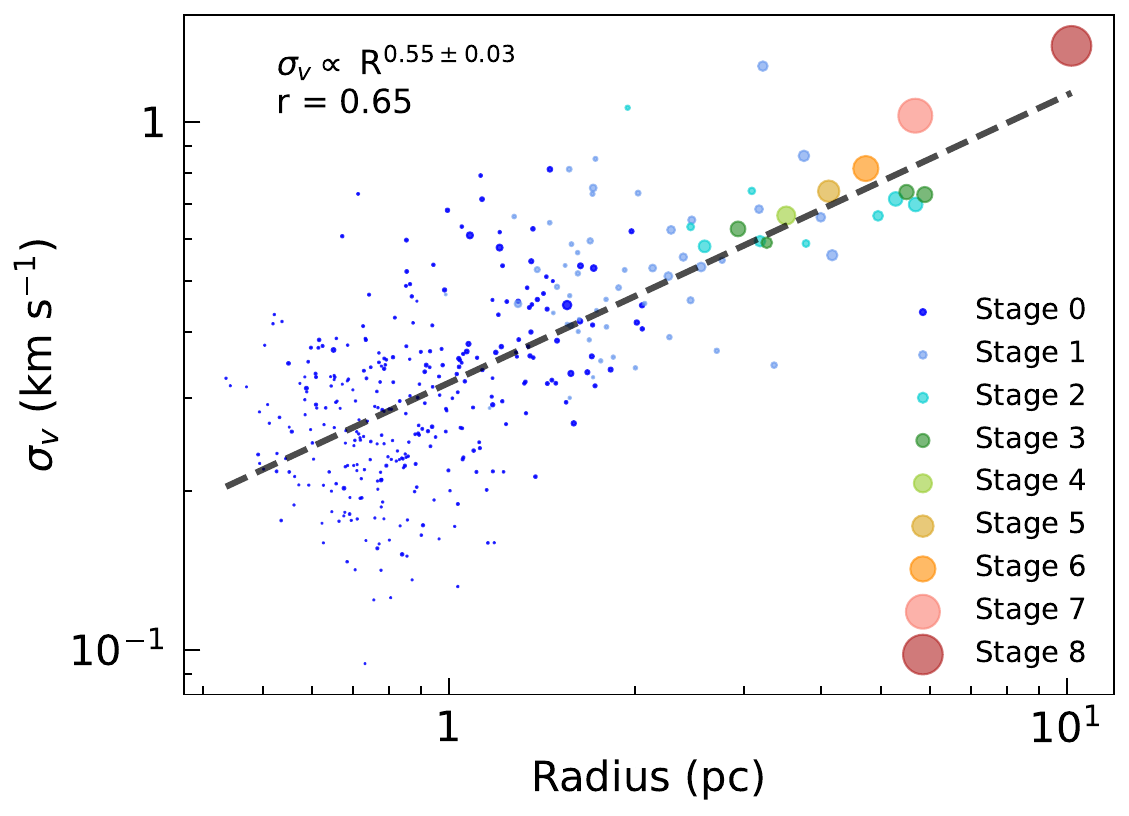}
   \\(a)
  \end{minipage}
  \begin{minipage}[t]{0.32\linewidth}
   \centering
      \includegraphics[width= \linewidth]{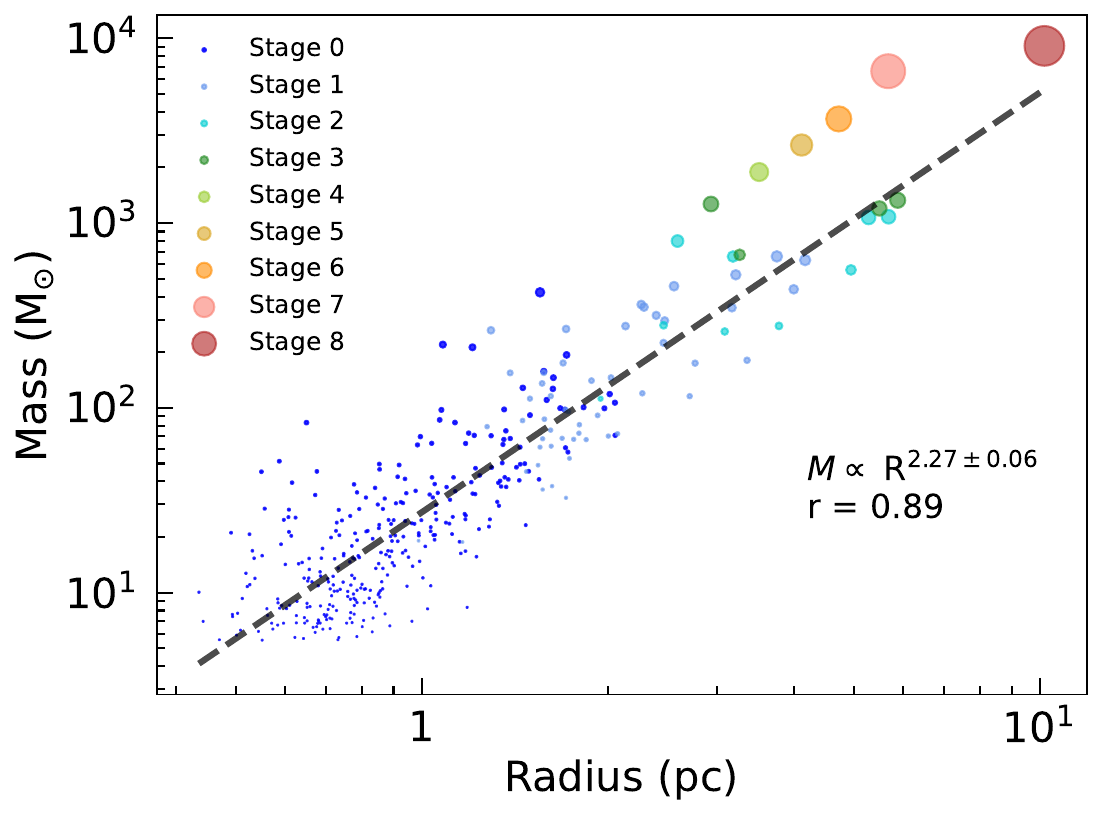}
   \\(b)
  \end{minipage}
  \begin{minipage}[t]{0.33\linewidth}
   \centering 
        \includegraphics[width= \linewidth]{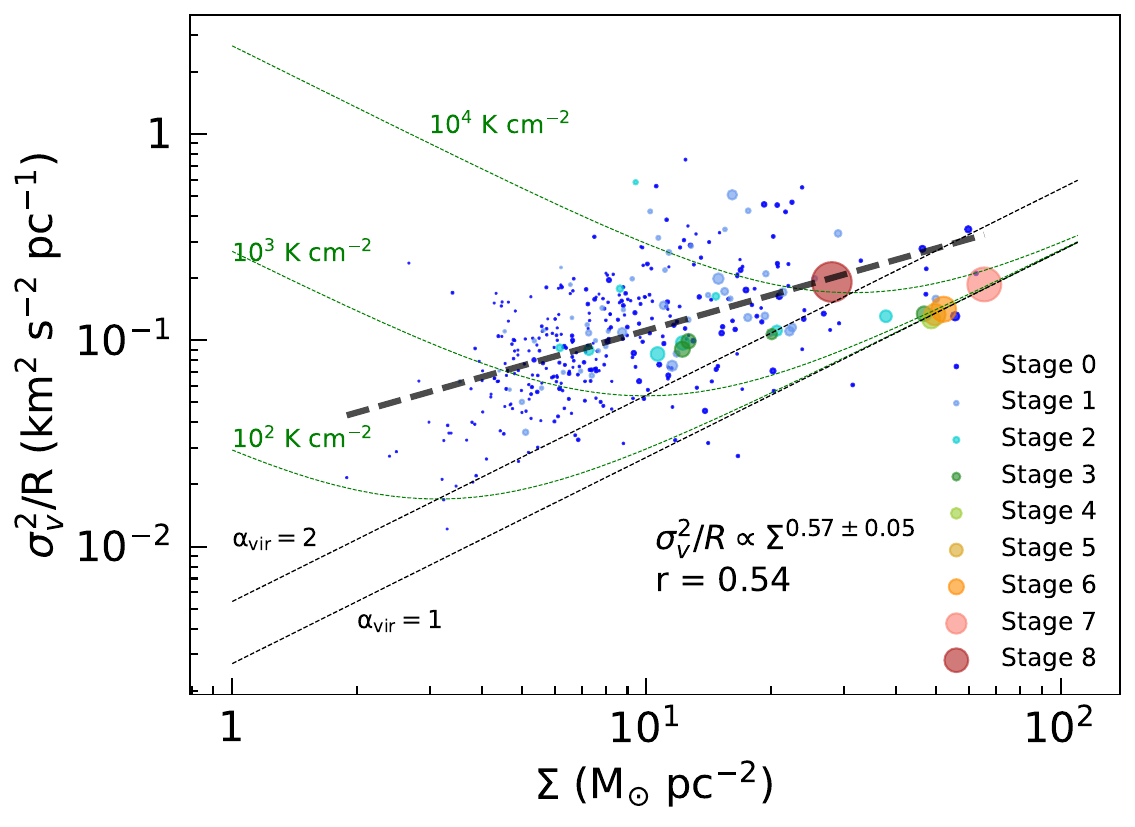}
   \\(c)
  \end{minipage}
\caption{Relationships between (a) velocity dispersion and radius, (b) mass and radius, and (c) parameter $\sigma_v^2/R$ and the surface density of the identified structures. Different colors in the figure represent structures in different stages. The marker size in the figure is proportional to the mass of the structures. } 
  \label{fig9}
 \end{figure*}
 
 \begin{figure*}
  \centering
  \begin{minipage}[h]{0.32\linewidth}
   \centering
   \includegraphics[width= \linewidth]{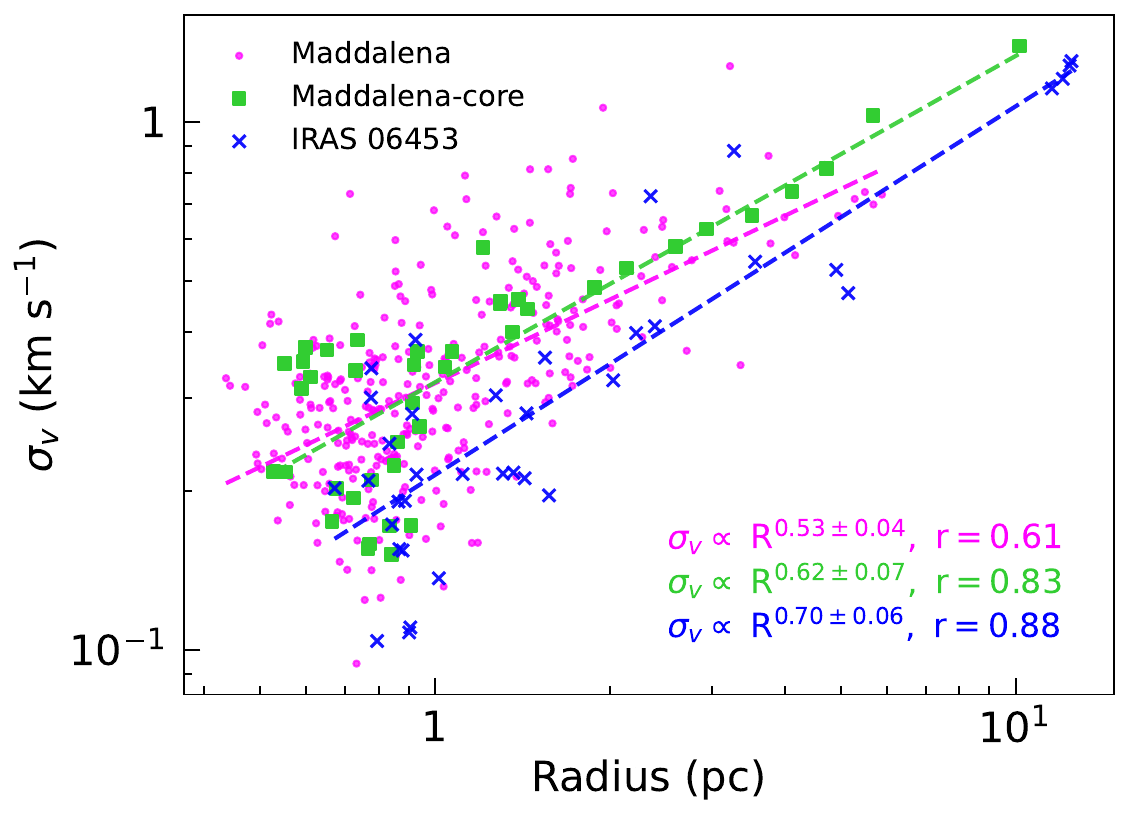}
   \\(a)
  \end{minipage}
  \begin{minipage}[h]{0.32\linewidth}
   \centering
      \includegraphics[width= \linewidth]{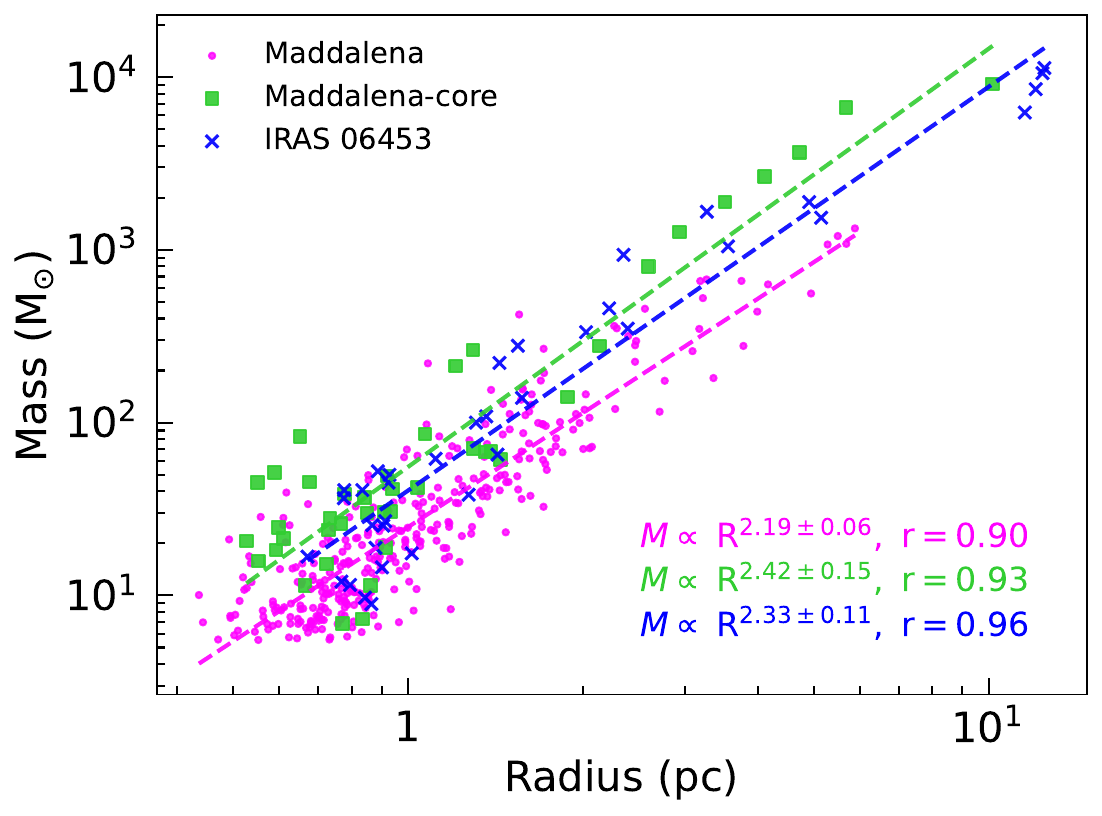}
   \\(b)
  \end{minipage}
  \begin{minipage}[h]{0.32\linewidth}
   \centering 
        \includegraphics[width= \linewidth]{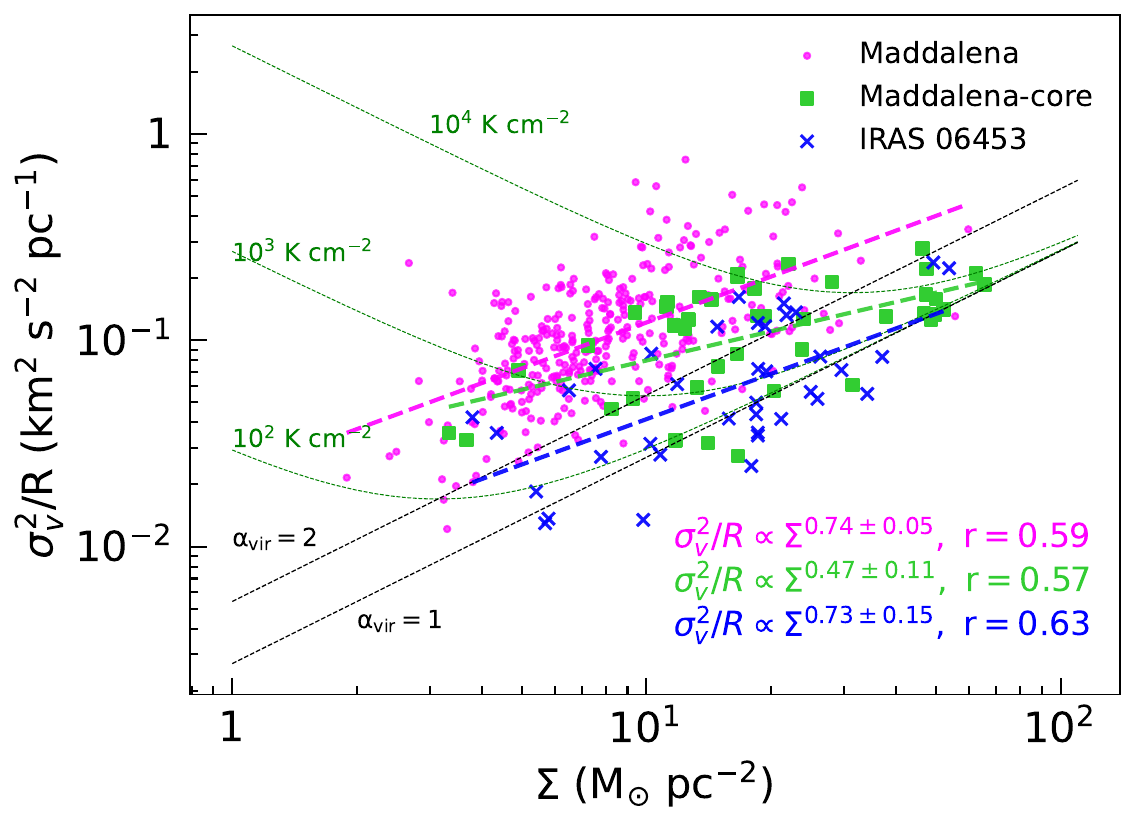}
   \\(c)
  \end{minipage}
  \caption{Similar to Figure \ref{fig9}, whereas the substructures in the IRAS 06453 star-forming region are included, and different colors represent different groups of structures classified in terms of star formation activity level. The groups are described in the text in Section \ref{sec3.2}.} 
  \label{fig10}
 \end{figure*}
 
\subsubsection{Relations between Different Physical Parameters} \label{sec3.3.2}
 
\citet{1981MNRAS.194..809L} found a power-law scaling relation between the line width and the size of molecular clouds, $\sigma_v\propto R^{0.38}$, which is attributed to the universal existence of turbulence in molecular clouds. Later, a power-law exponent of 0.5 in the $\sigma_v-R$ relation was obtained when a more homogeneous sample of molecular clouds was used \citep{1987ApJ...319..730S,2009ApJ...699.1092H}. A power-law $\sigma_v-R$ scaling behavior is frequently reported in the literature for molecular cores, clumps, and clouds, which span size from $\sim$0.01 to $\sim$50 pc, but all exhibit an exponent around 0.5 \citep{2001ApJ...562..348O,2006AJ....131.2921R,2008ApJ...686..948B, 2020MNRAS.499.2534K}. A steeper slope ($>$0.38) can result either from compressive turbulence \citep{2013MNRAS.436.3247K,2021ApJ...906L...4C} or from gravity-contraction-induced turbulence \citep{2011MNRAS.413.2935D,2016ApJ...824...41I}. We note that in these early studies the scaling relations are examined by measuring different entities (cores, clumps, or clouds) which may be located in quite different environments. In this work, the scaling relations are examined by measuring the structures of different levels in a single GMC. 

In Figure \ref{fig9} (a), the $\sigma_v$-R relation for the hierarchical structures identified in the Maddalena GMC is presented. As shown in Figure \ref{fig9} (a), structures in different stages with radius from $\sim$0.3 to $\sim$ 10 pc follow a power-law $\sigma_v-R$ scaling relation, $\sigma_v \propto R^{0.55\pm0.03}$, with a Pearson correlation coefficient between ln $\sigma_v$ and ln R of 0.65. The exponent is consistent with those in previous studies and seems invariant for structures in different stages, although there is a larger scatter for the structures at stage 0, i.e., the ``leaves''. For comparison, \cite{2001ApJ...551..852H} found that in the outer Galaxy the exponent is $\sim$0.5 for the clouds with radii above 9 pc. Figure \ref{fig9} (b) shows the mass-radius (m-r) relation, which can be well fitted by a power-law function with an exponent of 2.27, indicating that larger structures have higher mean column densities, which slightly conflicts with Larson's third law that molecular clouds have constant surface densities. 
 
In Figure \ref{fig9} (c), the relationship between $ \sigma_{v}^{2} / R$ and the surface density $\Sigma_{MC}$ of the hierarchical structures is presented. It reveals that the scaling of $\sigma_{v}^{2} / R$ with $\Sigma_{MC}$ follows a power-law relation with a power exponent of 0.57, which differs from the findings by \citet{2009ApJ...699.1092H}. The black dashed lines in Figure \ref{fig9} (c) show the relationships corresponding to simple virial equilibrium(SVE; $\alpha =1$) and marginally bound clouds($\alpha=2$), respectively. When external pressure is considered, the equilibrium solutions for molecular structures can be expressed with the green dotted lines in Figure \ref{fig9} (c) \citep{2011MNRAS.416..710F}. As seen in Figure \ref{fig9} (c), the majority of the structures in the Maddalena GMC distribute above the SVE and the marginally gravitationally bound lines. The expected external pressure which helps to bound these structures and prevent rapid dissipation should be within the range of $10^2$ to $10^4$ K cm$^{-2}$. Interestingly, structures of larger scales, which usually are at higher stages, show a tendency to be more gravitationally bound, while most of the structures on small to medium stages show no discernible differences in the $\sigma_v^2/R\propto\Sigma$ relation.

For the comparison of scaling relations between star forming region and quiescent region, we present the scaling relations of the Maddalena, Maddalena-core, and IRAS 06453 in Figure \ref{fig10} separately with different colors. The substructures within the Maddalena core and quiescent regions display similar $\sigma_v-R$ relations, with exponents around $\sim$0.6, while the substructures in the IRAS 06453 region have lower velocity dispersions and exhibit a steeper power-law scaling with an exponent of around 0.7. No obvious difference in the $\sigma_v-R$ scaling relation is seen between star forming and quiescent regions. Figure \ref{fig10}(b) reveals that substructures in the three groups follow quite similar M-R relationships with power-law exponents within 2.3-2.4. However, the substructures in Maddalena-core and IRAS 06453 are systematically more massive than those in the quiescent regions, indicating higher column densities for molecular cloud structures which host star formation. Consistently, in Figure \ref{fig10} (c) structures within the two star-forming regions exhibit closer alignment with the SVE and marginally gravitationally bound lines than the structures in the quiescent region. Around half of the substructures in Maddalena-core region are distributed below the $\alpha_{vir} = 2$ line. In contrast, structures in quiescent region are hardly self-gravitating. The above results are consistent with the classic picture that star formation activities usually take place in dense self-gravitating regions.
 
\begin{figure*}
\centering
\begin{minipage}[h]{0.45\linewidth}
  \centering
  \subfigure[]{
  \raisebox{0.05\height}{\includegraphics[width= \linewidth]{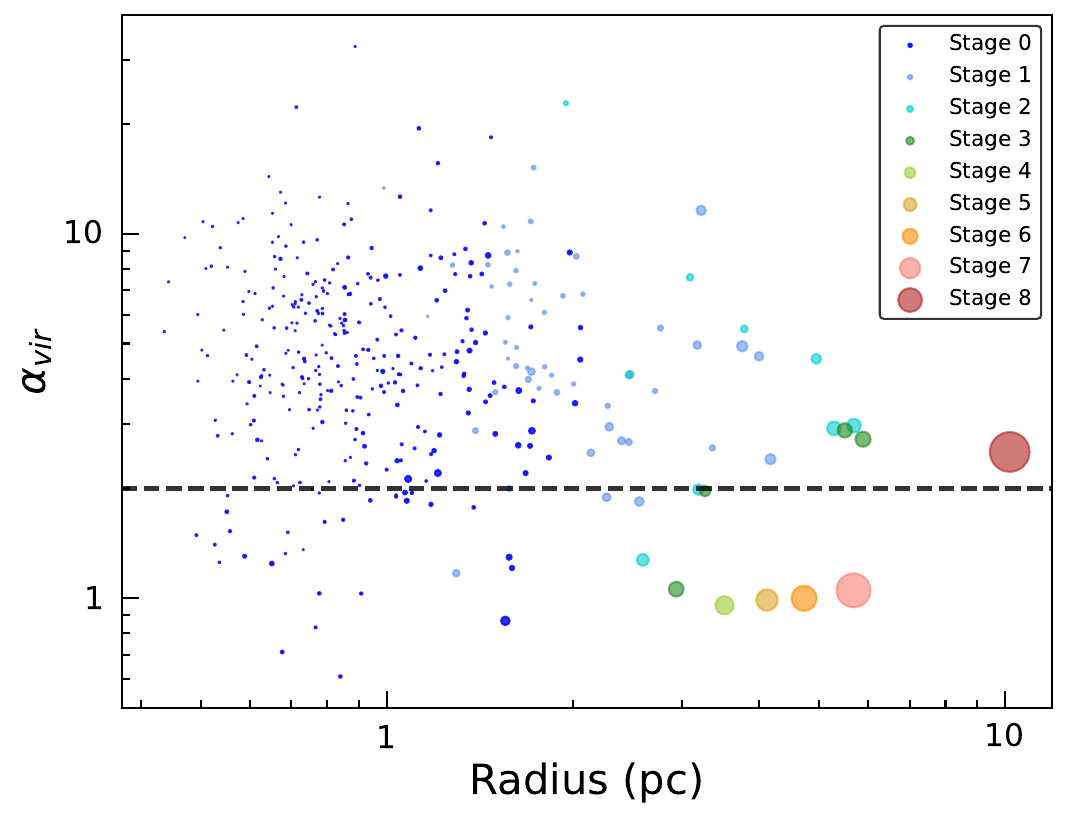}}}
\end{minipage}
\begin{minipage}[h]{0.45\linewidth}
  \centering 
  \subfigure[]{
  
  \raisebox{0.05\height}{\includegraphics[width= \linewidth]{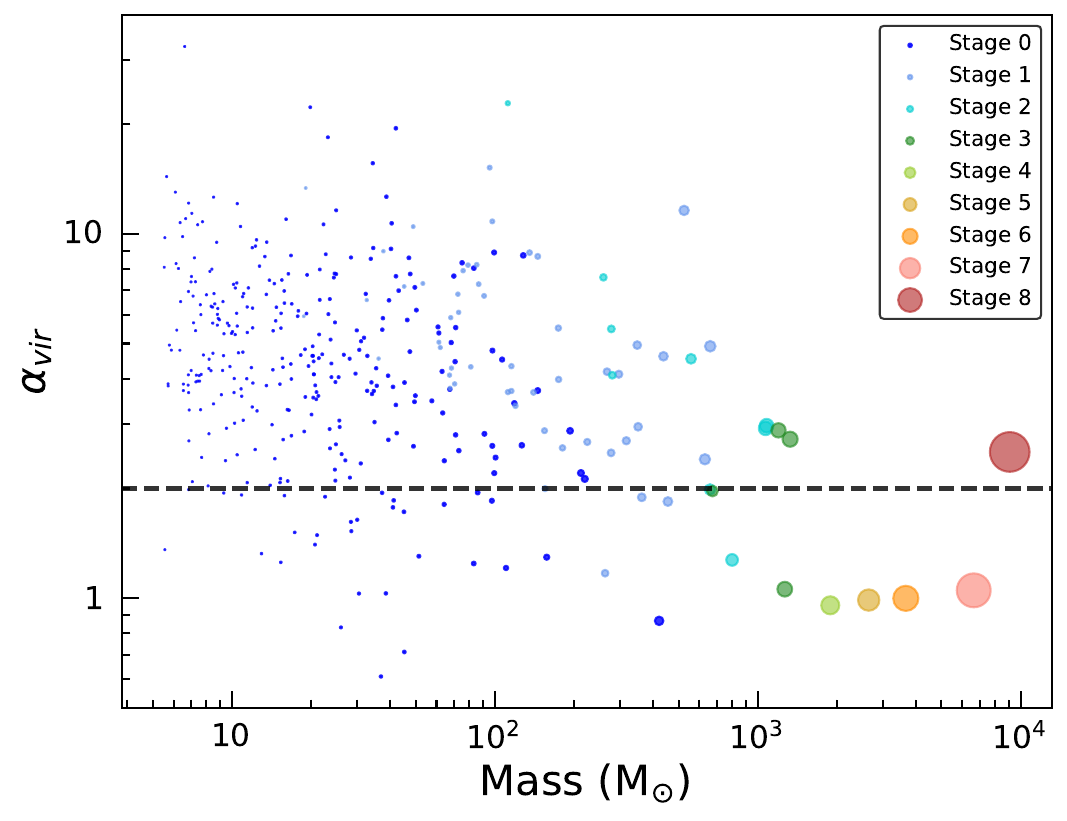}}}
\end{minipage}
\begin{minipage}[h]{0.45\linewidth}
  \centering
  \subfigure[]{

  \raisebox{0.05\height}{\includegraphics[width= \linewidth]{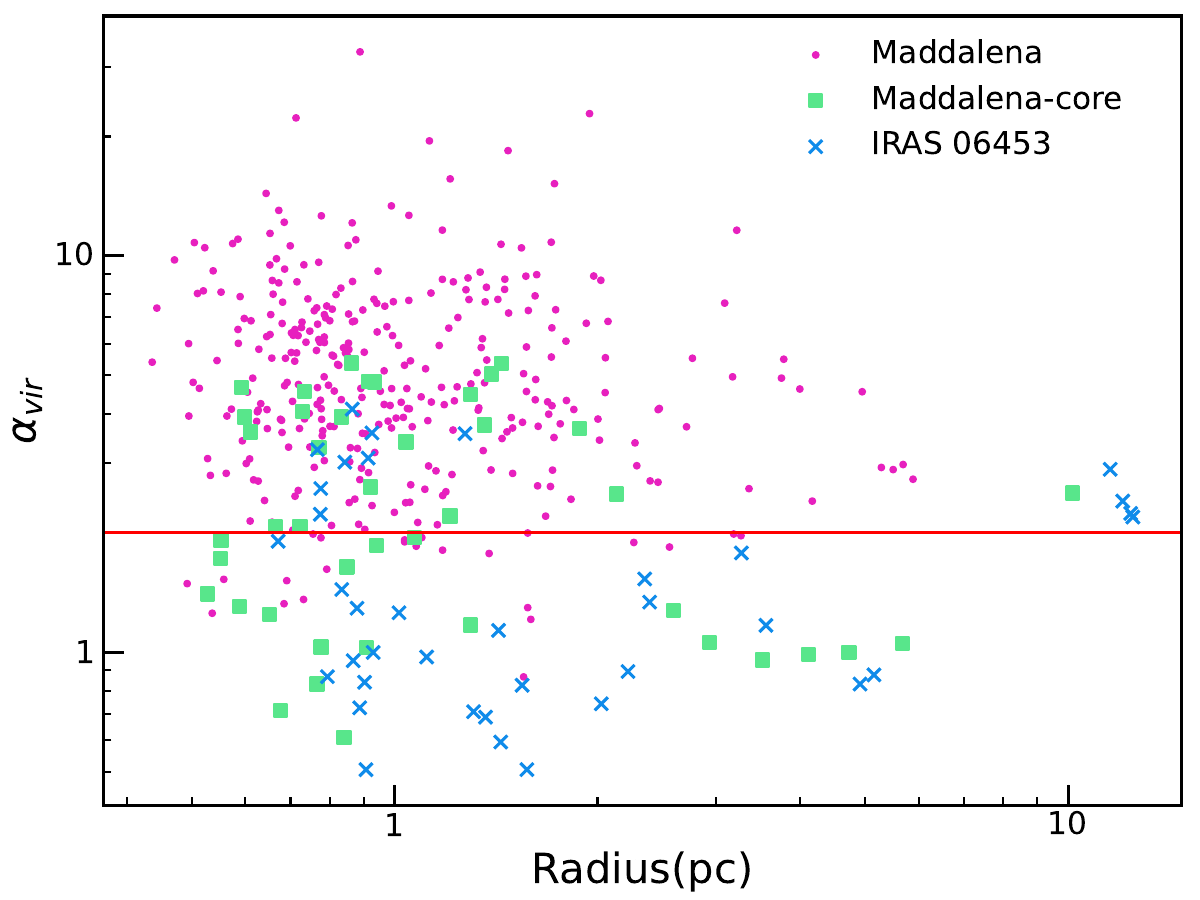}}}

\end{minipage}
  \begin{minipage}[h]{0.45\linewidth}
  \centering 
  \subfigure[]{

  \raisebox{0.05\height}{\includegraphics[width= \linewidth]{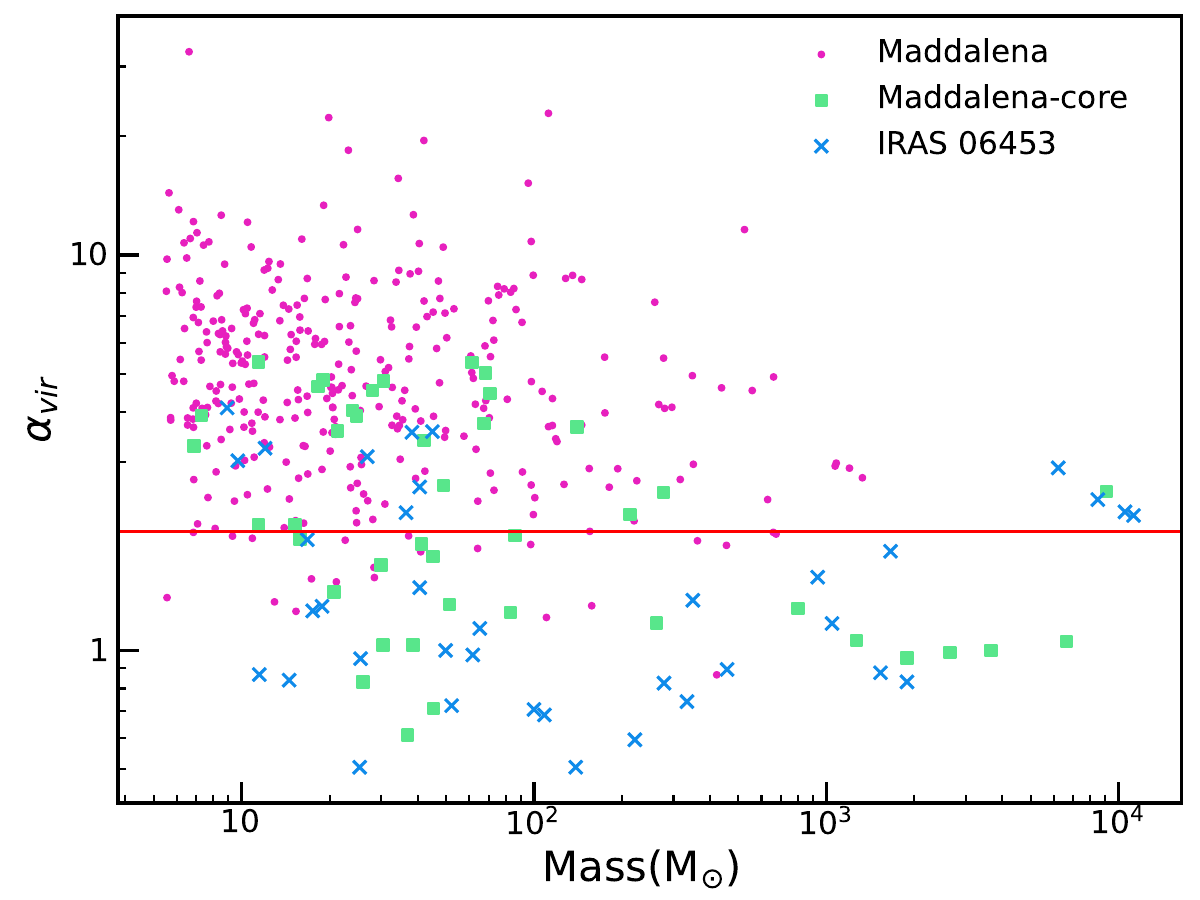}}}
\end{minipage}
\caption{(a) Relationship between virial parameter and radius of the structures. Different colors indicate different stages, stage = 0 represents the leaves in the tree diagram, stage = 1 means it contains one layer of substructure, stage = 2 contains two layers of substructure, and so on. (b) Same as panel (a), but for the relationship between virial parameters and mass. (c) Same as panel (a), but with colors representing different groups of structures categorized in terms of star formation activity level. (d) Same as panel (c), but for the relationship between virial parameter and mass. The grey dashed lines in panels (a) and (b) and the red lines in panels (c) and (d) correspond to the critical virial parameter $\alpha_{vir}=2$.} 
\label{fig11}
\end{figure*}

\section{Discussion}\label{sec4}
\subsection{Self-gravitation Across Scale}
Figures \ref{fig11}(a) and \ref{fig11}(b) present the relationship between radius/mass and the virial parameter of the structures identified in the region. The size of the data points is proportional to the mass of the structures, and various colors are used to indicate different stages. For structures other than at stage = 0, we can see a trend that the virial parameter decreases with increasing radius/mass, while for the structures at stage = 0, this trend is not evident. For comparison, we present the structures belonging to Maddalena, Maddalena-core, and IRAS 06453 separately in Figures \ref{fig11}(c) and \ref{fig11}(d). A noticeable distinction is observed between the non-star-forming region and star-forming region. For Maddalena, the majority of the structures are not gravitationally bound (i.e., $\alpha_{vir}>2$), while for Maddalena-core and IRAS 06453, most of the structures are gravitationally bound (i.e., $\alpha_{vir}<2$). 
 
 \begin{figure*}
  \centering
   \begin{minipage}[t]{0.47\linewidth}
   \centering 
   \subfigure[]{
  \includegraphics[width= \linewidth]{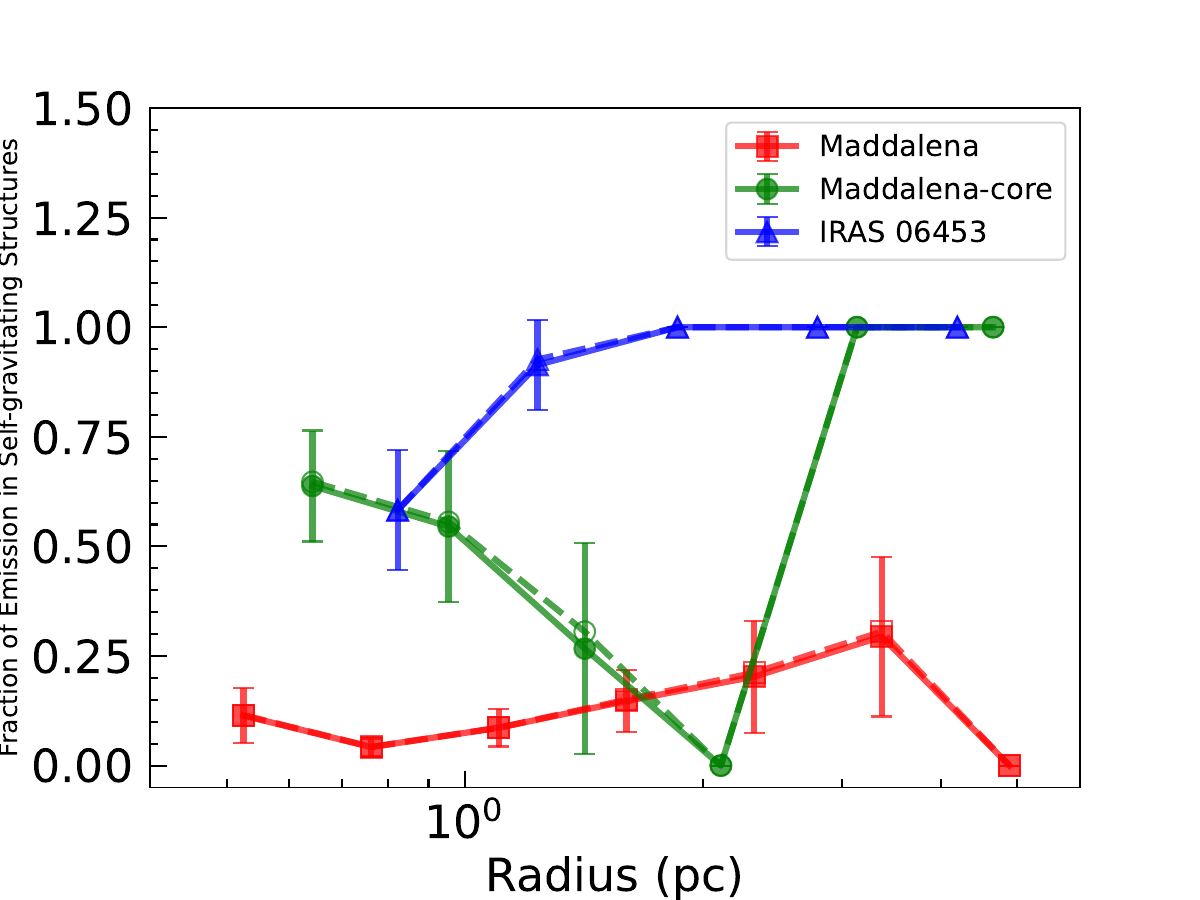}}
  \end{minipage}
   \begin{minipage}[t]{0.47\linewidth}
   \centering 
   \subfigure[]{
   \includegraphics[width= \linewidth]{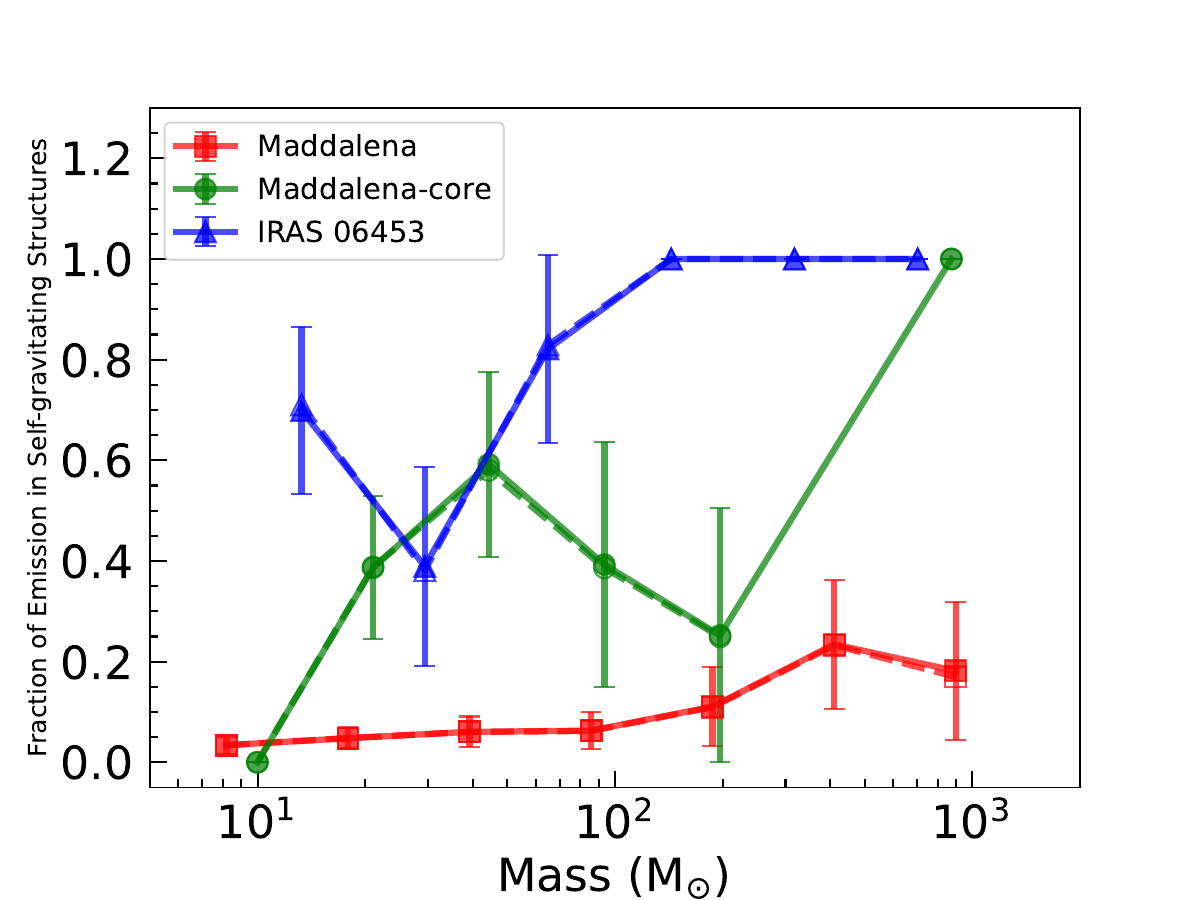}}
  \end{minipage}
  \caption{Fraction of self-gravitating emission as a function of (a) radius and (b) mass of the structures in the Maddalena, Maddalena-core, and IRAS 06453 regions. Filled symbols show the averages from the bootstrap resampling and are connected with solid lines. Hollow symbols, which almost coincide with the the filled symbols,  present the results without bootstrap resampling, and are connected with dashed lines.}
  \label{fig12}
 \end{figure*}
 
 \begin{figure*}[htb!]
  \gridline{\fig{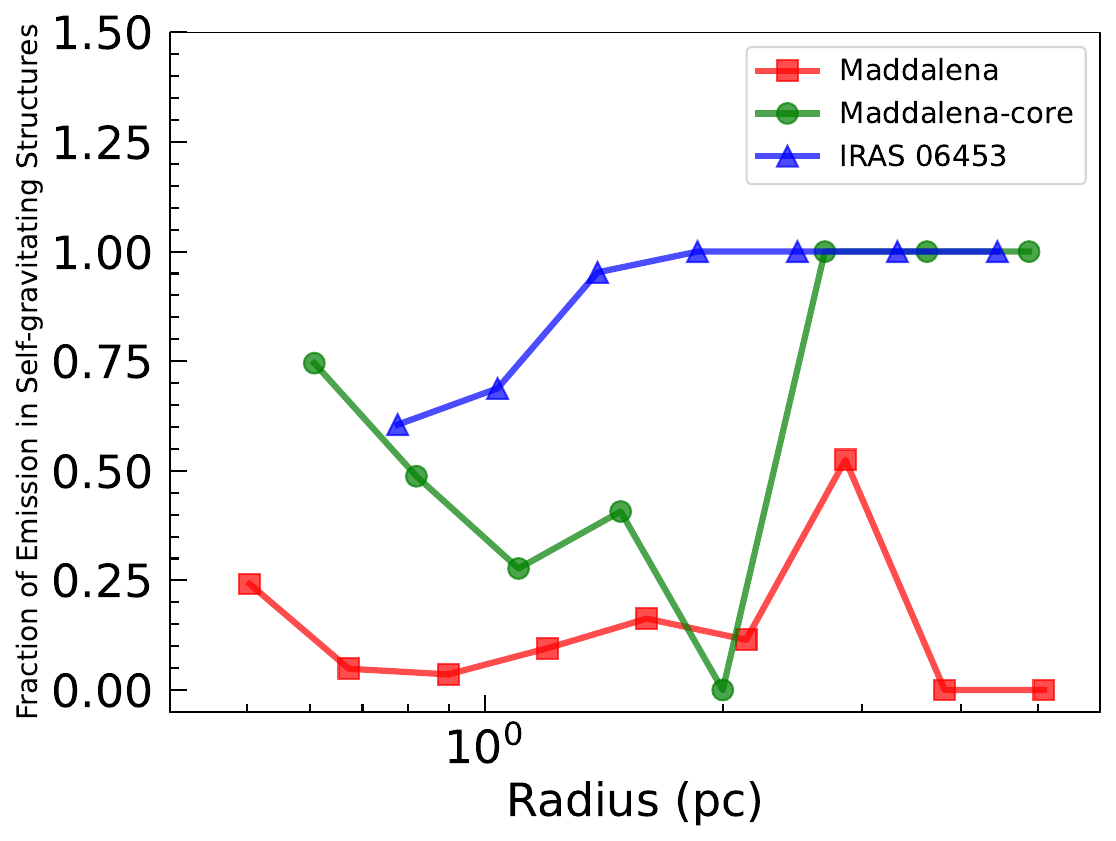}{0.25\textwidth}{(a)}
  \fig{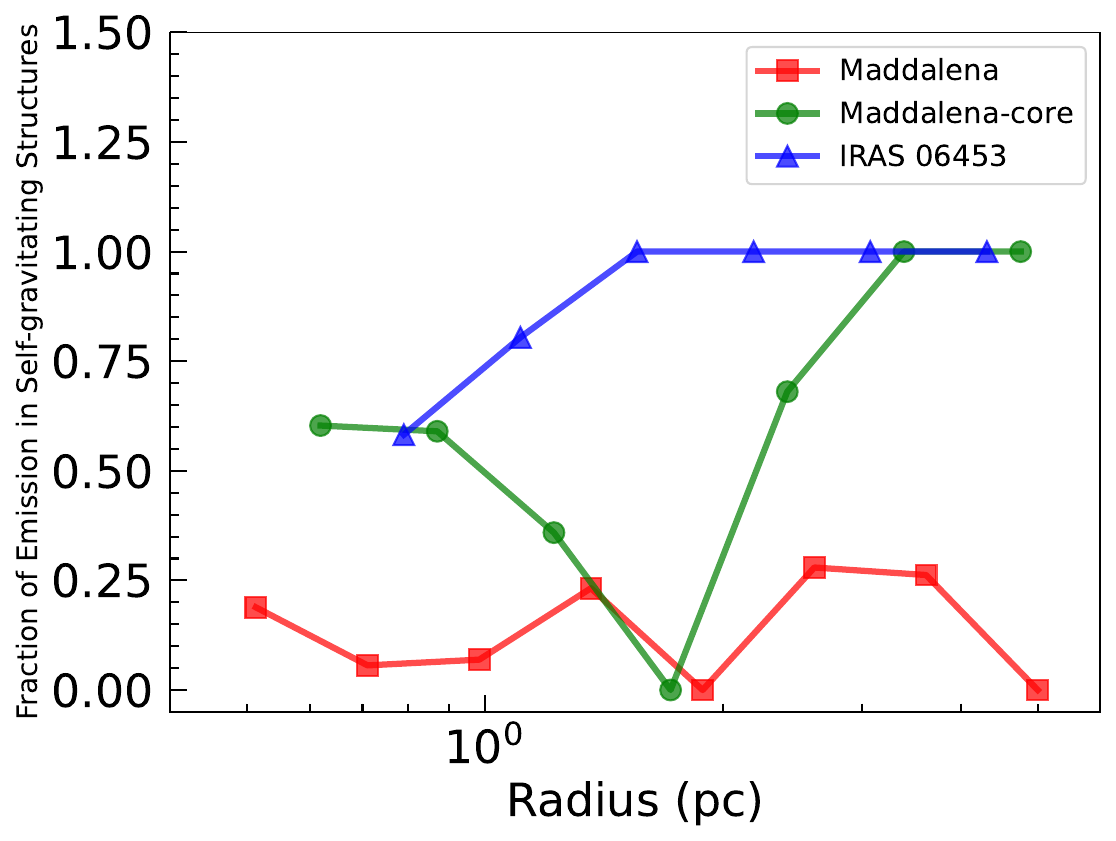}{0.25\textwidth}{(b)} 
  \fig{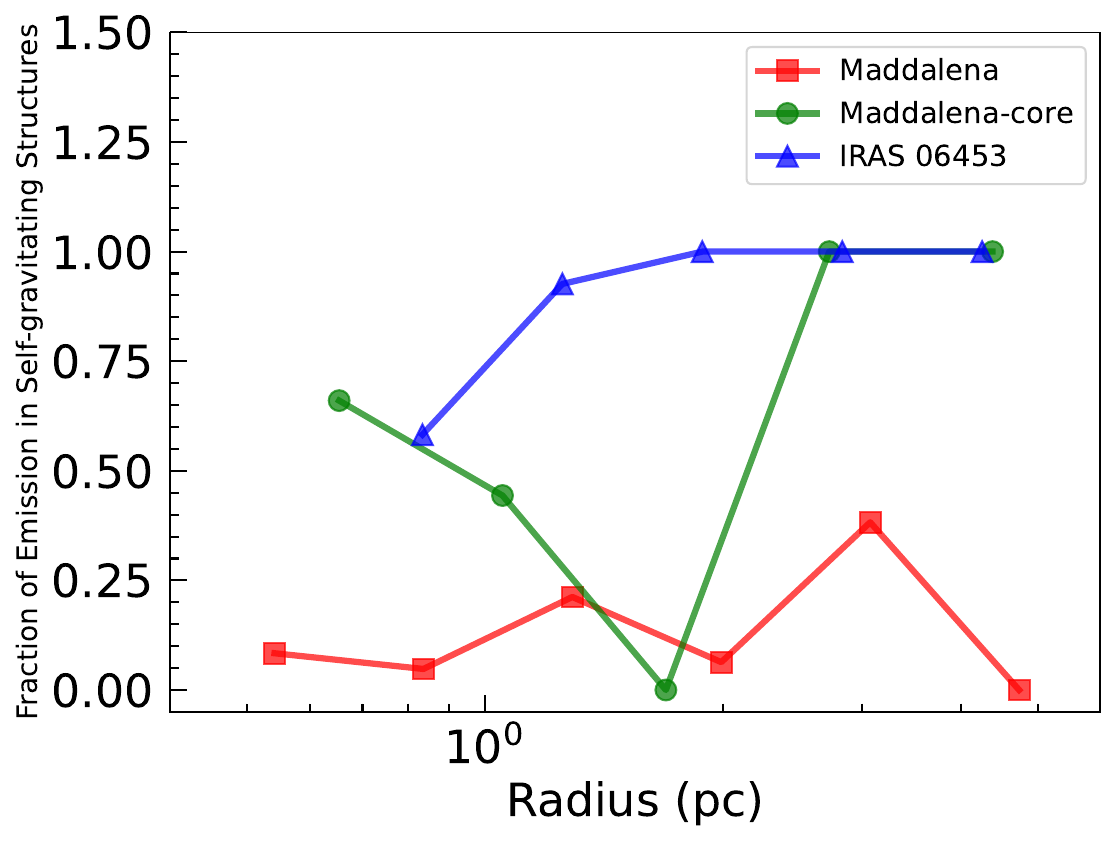}{0.25\textwidth}{(c)}
  \fig{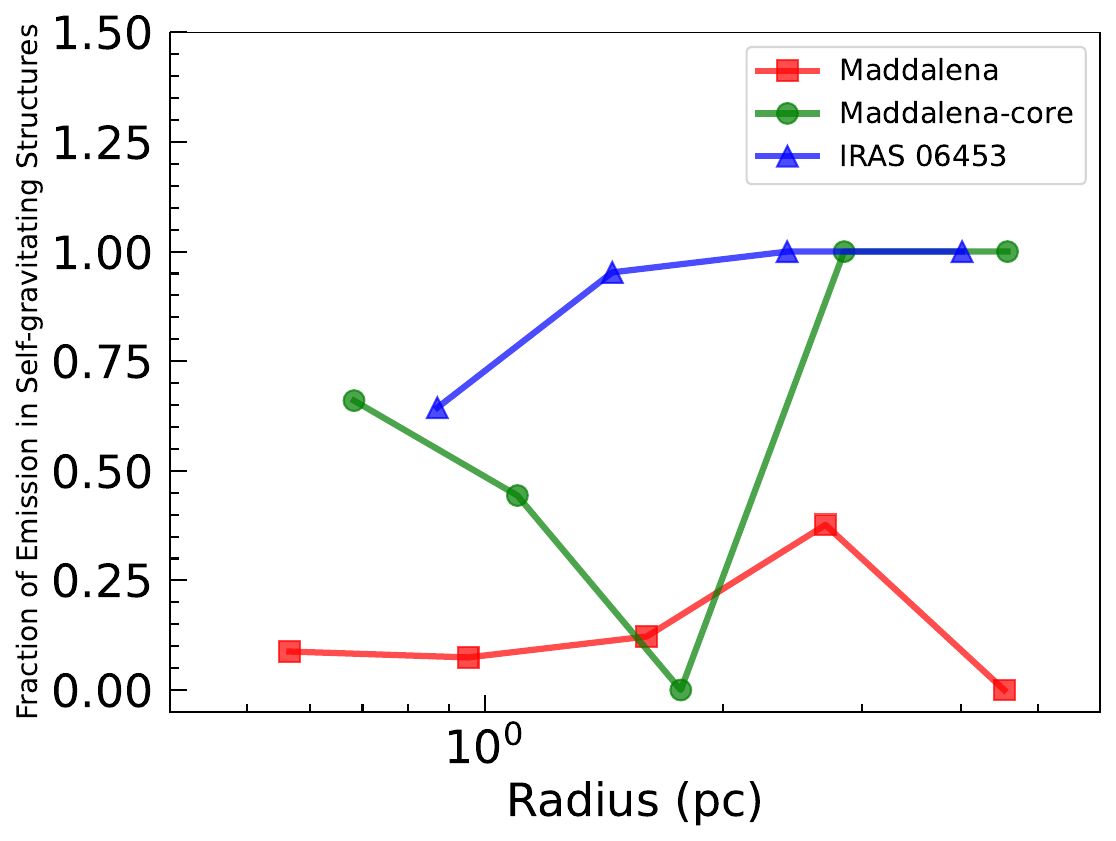}{0.25\textwidth}{(d)}}
  \gridline{\fig{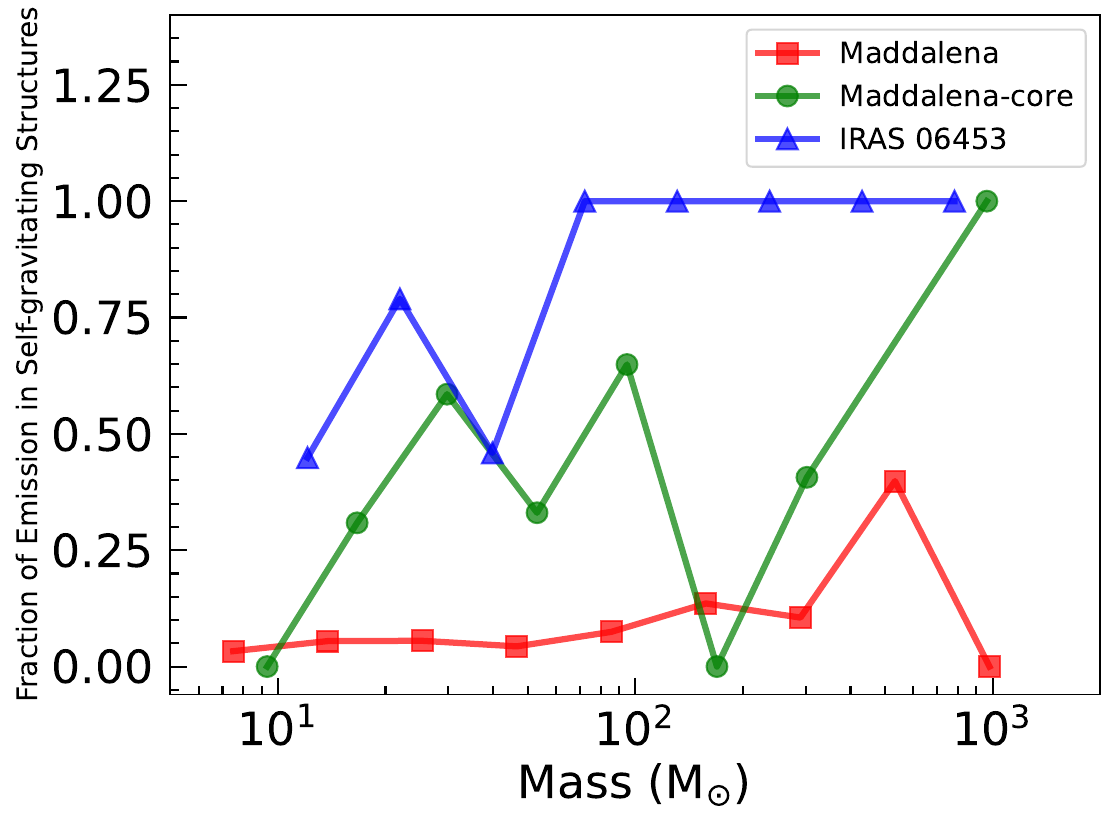}{0.25\textwidth}{(e)}
  \fig{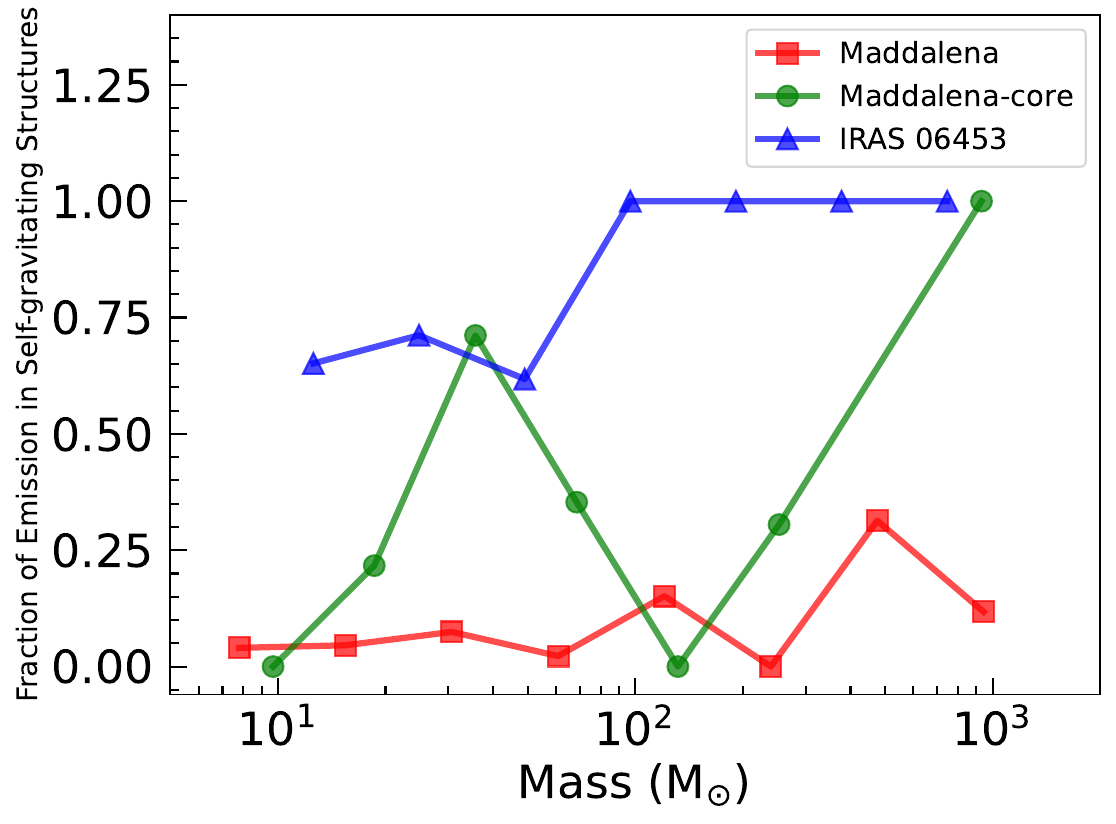}{0.25\textwidth}{(f)}
  \fig{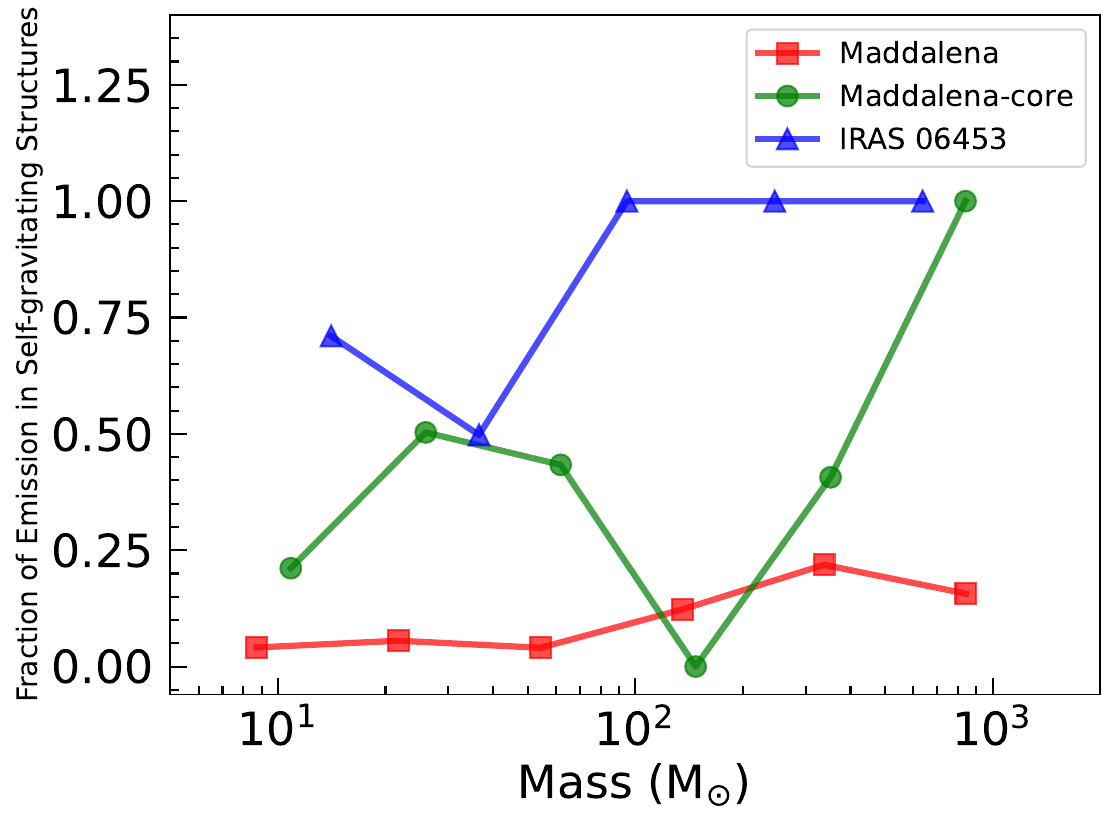}{0.25\textwidth}{(g)}
  \fig{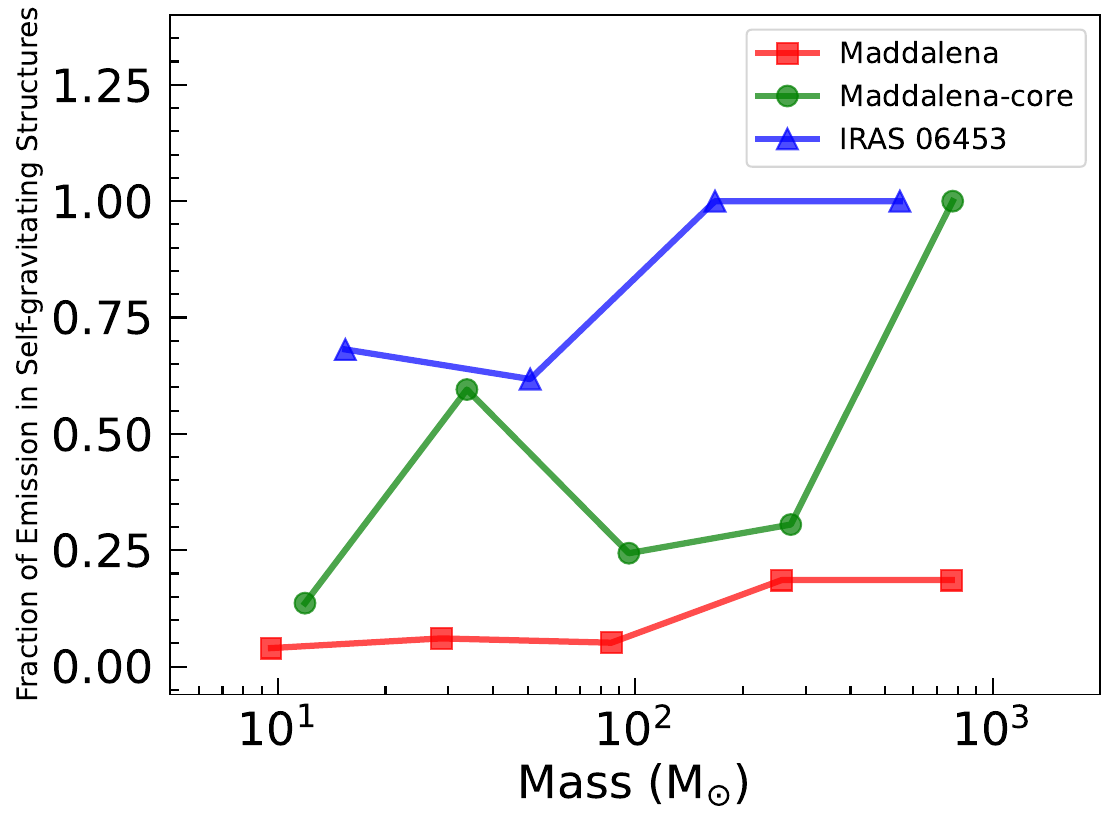}{0.25\textwidth}{(h)}}
  \caption{Same as Figure \ref{fig12}, but using different numbers of bins. Panels (a)-(d) correspond to using 9, 8, 6, and 5 bins, respectively, for the fraction-radius relation, while panels (e)-(h) correspond to using 9, 8, 6, and 5 bins, respectively, for the fraction-mass relation.}
  \label{fig_nbins}
\end{figure*}
 
Figure \ref{fig12} shows the variation of the fraction of self-gravitating structures as a function of radius and mass in the Maddalena, Maddalena-core, and IRAS 06453 regions. The fraction at a specific radius/mass is defined as the luminosity fraction that the self-gravitating structures account for the total luminosity of the structures within a radius/mass bin. We divide the ranges of the radius ($\sim$ 0.4-6 pc) and the mass ($\sim$ 5-$1.5 \times 10^{3}$ M$_{\odot}$) of the structures in Maddalena into seven bins of equal size on the logarithmic scale in Figures \ref{fig12}(a) and \ref{fig12}(b), respectively. The numbers of bins in Figure \ref{fig12} for the Maddalena core and the IRAS 06453 regions are adjusted to maintain the bin sizes comparable to that for the Maddalena region. Results when adopting more or fewer bins are presented in Figure \ref{fig_nbins}.

The uncertainties for the data points show in Figures \ref{fig12} and \ref{fig_nbins} are hard to estimate, as shown by the formula A1-A6 list in the Appendix which are used in our calculation of the virial parameter. The error bars in Figure \ref{fig12} are random errors that we derived with the bootstrap method. For this purpose, we resampled the identified structures 10, 000 times.

In both panels of Figure \ref{fig12}, the fraction of gravitating structures in the Maddalena quiescent region is consistently lower across almost all scales ($\sim$0.5-5 pc) than that for the two star-forming regions. The fraction of self-gravitating structures in the Maddalena quiescent region, as shown by the red line in Figure \ref{fig12} (a), decreases from $13\%$ to $\sim 5\%$ as the radius increases from 0.5 to 0.8 pc. Then, it increases gradually from $10\%$ to $20\%-30\%$ as the radius extends from 0.8 to 4 pc. Similar behavior is observed in the Maddalena-core region, where the fraction of self-gravitating structures initially decreases from $65\%$ to zero as the radius increases from 0.6 to 2 pc and then increases sharply from zero to $100\%$ as the radius extends from $\sim$2 pc to 3-5 pc. The IRAS 06453 cloud differs from the aforementioned two regions in that its fraction of self-gravitating structures increases from 55\% to 100\% as the radius increases from approximately 0.8 to approximately 2 pc, and then remains at 100\% from 2 to 4 pc. The fraction of self-gravitating structures as a function of the structure mass is presented in Figure \ref{fig12} (b). For the Maddalena region, the behavior of the fraction with the structure mass is similar to that with structure radius. The ``Maddalena-core" curve, represented by the green line, begins at zero near a mass of 10 M$_{\odot}$, then rises to approximately 60$\%$ at around 40 M$_{\odot}$ and shows a decrease past this point, exhibiting a dip at about 200 M$_{\odot}$. Subsequently, it increases from 25$\%$ to 100$\%$ as the mass increases from 200 M$_{\odot}$ to 900 M$_{\odot}$. For the IRAS 06453 region, the overall fraction of gravitationally bound structures is high ($>60\%$), except that there is a significant reduction at around 30 M$_{\odot}$. From Figure \ref{fig_nbins}, we can see that the above discussed trends do not change when different bin sizes are used.

 
In this work, we report that self-gravity is important on spatial scales from $\sim$0.8 to $\sim$4 pc for the IRAS 06453 star-forming region. In the Maddalena-core region, self-gravity is important at scales smaller than $\sim$1 pc or greater than $\sim$2 pc. \citet{2009Natur.457...63G} found that self-gravity plays an important role at all scales from 0.1 to $\sim$ 1 pc in the star-forming cloud L1448. However, we note that in the Maddalena quiescent region, self-gravitating structures only account for less than 25$\%$ of structures on all possible scales below $\sim$6 pc, indicating that self-gravity is hardly a dominant factor for structures in this region. 
 
\subsection{Comparison with Numerical Simulations}
 
\begin{figure}
  \includegraphics[width=\columnwidth]{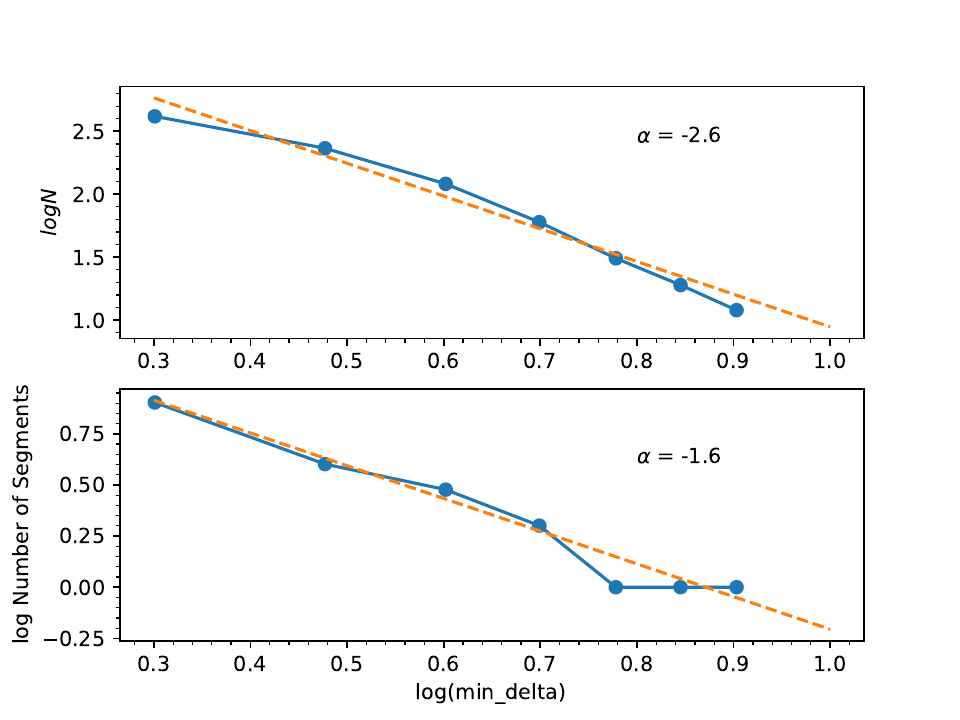}
    \caption{Top: relationship between total numbers of structures (leaves and branches) and $min\_delta$. Bottom: relationship between number of levels of the biggest branch and $min\_delta$.}
     \label{fig13}
\end{figure}

\cite{2013ApJ...770..141B} found significant correlations between the hierarchical complexity of molecular clouds and their sonic and Alfv\'{e}nic Mach numbers, as well as self-gravity. Their study reveals that simulations including self-gravity, stronger magnetic fields, and higher sonic Mach numbers can generate increased numbers and levels of substructures. They found power-law relationships between the total number of substructures, $N_{\text{tot}}$, the maximum number of levels, $N_{\text{level}}$, and the input parameter, $min\_delta$. With different sonic and Alfv\'{e}nic Mach numbers, the $N_{\text{tot}}$-$min\_delta$ and $N_{level}$-$min\_delta$ relations exhibit different power-law indices. In this work, we examine the relationships of $N_{\text{tot}}$ and $N_{\text{level}}$ versus the parameter $min\_delta$ using observational data. In the generation of dendrograms with different $min\_delta$, the parameter $branch\_delta$ is always set to be one $\sigma_{13,\rm RMS}$ smaller than $min\_delta$. The $ N_{\text{tot}}$-$min\_delta$ and $ N_{\text{level}}$-$min\_delta$ relations derived from the $^{13}$CO data of the Maddalena GMC are given in Figure \ref{fig13}. We can see that the two relations follow power-law with indices of $-$2.6 and $-$1.6, respectively. Interestingly, these power-law indices are comparable to those obtained in the run of numerical simulations with a sonic Mach number of 3 and an Alfv\'{e}nic Mach number of 0.7, as demonstrated in figure 4 of \cite{2013ApJ...770..141B}. As discussed in Section \ref{sec3.3.1} and shown in Figure \ref{fig8}, the median sonic Mach number observed in the Maddalena GMC is around 3.5. Our case study shows that the $ N_{\text{tot}}$-$min\_delta$ and $ N_{\text{level}}$-$min\_delta$ relationships can be used to reveal the strengths of turbulence and magnetization of GMCs.   
 
\section{Conclusion}\label{sec5}

In this study, we explore the structure, turbulence, and self-gravity of the Maddalena GMC using the high-sensitivity data of multiple lines of CO and its isotopologues of the region from the MWISP survey. Spatial distributions of integrated intensity, velocity, velocity dispersion, excitation temperature, and column density maps of the region are presented. We decompose the $^{13}$CO PPV datacube into hierarchical substructures using a modified dendrogram algorithm and conduct an analysis on the statistical properties of the identified substructures. The dendrogram algorithm interprets observation data in a hierarchical manner, which allows to investigate variations of physical properties of molecular gas on various scales within one single GMC. In this work, we focus on understanding the role of self-gravity across scale in the Maddalena GMC. The main results are summarized as follows.

\begin{enumerate}  
\item  The majority of the CO ($J=1-0$) emission from the Maddalena GMC is within the velocity range of 15 to 38 km s$^{-1}$. A velocity gradient of about $\sim$0.08 km s$^{-1}$ pc$^{-1}$ is observed in the GMC along the direction from southwest to northeast. This GMC is notably cold, with temperatures below 10 K in most of the region, and a peak excitation temperature of only 16.5 K. The entire GMC has quite low H$_2$ column densities, ranging from 1.24$\times$10$^{18}$ to 1.27$\times$10$^{22}$ cm$^{-2}$, with a median value of 1.21$\times$10$^{21}$ cm$^{-2}$. These results are consistent to previous studies \citep{1994ApJ...432..167L}.

\item  Utilizing the modified dendrogram algorithm, we have identified 377 substructures within the Maddalena GMC, excluding structures associated with the two satellite IRAS sources located within the observed region, IRAS 06453 and IRAS 06522. 
Among the identified substructures, 197 are trunks, 73$\%$ of which are monadic and 27$\%$ of which contains at least 2 substructures, i.e., hierarchical. The radius and mass of the substructures follow power-law distributions with indices of $-$2.59 and $-$1.64, respectively. Compared with the hierarchical substructures, monadic substructures are generally much smaller and have much smaller mass, with a typical size of a few tenths of a parsec and a typical mass of tens solar masses. The monadic and hierarchical substructures both show mass-radius scaling relations that have similar power-law exponents. However, the hierarchical substructures have higher surface densities than the monadic substructures. These behaviors are similar to the results derived from traditional segmentation algorithms \citep{2001ApJ...551..852H, 2016ApJ...822...52R,2017ApJ...834...57M}. Most of the substructures of the Maddalena GMC are moderately supersonic, with a median sonic Mach number around $\sim$3.5.

\item  We conducted an analysis of the variation of virial parameters versus the radius/mass of the substructures. For hierarchical substructures, we find a decreasing trend in the virial parameter as the radius or mass of the substructure increases. Notably, the majority of the substructures in the quiescent Maddalena region are not gravitationally bound ($\alpha_{vir}>2$), while most of the substructures in the star-forming regions, Maddalena-core and IRAS 06453, are gravitationally bound ($\alpha_{vir}<2$). Using the fraction of the self-gravitating substructures as an indicator for the dominance of self-gravity, we find that self-gravity plays an important role across the whole range of the scale, 0.8-4 pc, of the substructures identified in the IRAS 06453 star-forming region, while for substructures identified in the non-star-forming region, self-gravity is not an important factor on all possible scales below 5 pc. 

\end{enumerate}

\begin{acknowledgments}
We thank the anonymous referee for the constructive comments that help to improve this manuscript.
This research made use of the data from the Milky Way Imaging Scroll Painting (MWISP) project, which is a multi-line survey in $^{12}$CO/$^{13}$CO/C$^{18}$O along the northern galactic plane with PMO-13.7 m telescope. We are grateful to all the members of the MWISP working group, particularly the staff members at PMO-13.7 m telescope, for their long-term support. MWISP was sponsored by National Key R$\&$D Program of China with grants 2023YFA1608000 $\&$ 2017YFA0402701 and by CAS Key Research Program of Frontier Sciences with grant QYZDJ-SSW-SLH047. H.W. acknowledges the support of NSFC grant 11973091. Y.M. acknowledges the support of NSFC grants 12303033 and 11973090. M. Z. acknowledges the support of NSFC grant 12073079. This research made use of astrodendro, a Python package to compute dendrograms of Astronomical data (\url{http://www.dendrograms.org/}).

\end{acknowledgments}

%

\vspace{5mm}





\appendix
\section{Calculation of Physical Parameters}
\label{sec:maths} 

The physical properties of the molecular cloud discussed in the text is calculated following the equations below. First, we assume that the molecular cloud is in LTE condition and the background source is the 2.7 K cosmic microwave background radiation. For the optically thick spectral lines, we can obtain the excitation temperature as \citep{1998AJ....116..336N,2000Obs...120..289R},

\begin{equation}
    T_{\rm ex}=\frac{5.53}{\ln{(1+\frac{5.53}{T_{peak}+0.819}})}.
	\label{eqA1}
\end{equation}
where $T_{peak}$ is the peak temperature of the $^{12}CO(1-0)$ spectra. We assume that the excitation temperatures for the $^{12}$CO and $^{13}$CO emission are the same, then we can estimate the optical depth of the $^{13}CO(1-0)$ spectra. The equation is given by \cite{2010ApJ...721..686P},
\begin{equation}
    \tau(^{13}CO)=-\ln({{1-\frac{T_{mb}(^{13}CO)}{5.29[J_1(T_{\rm ex})-0.164]}}})
	\label{eqA2}
\end{equation}
where $J_{1}\left(T_{\mathrm{ex}}\right)=1 /\left[\exp \left(5.29 / T_{\mathrm{ex}}\right)-1\right]$.
The total column density of $^{13}$CO can be calculated from the optical depth and the excitation temperature using the following formula \citep{1991ApJ...374..540G,1997ApJ...476..781B}:
\begin{equation}
    N(^{13}CO)=2.42\times10^{14}\frac{T_{\rm ex}+0.88}{1-exp(-5.29/T_{\rm ex})}\cdot\int\tau(^{13}CO)d\nu
    \label{eqA3}
\end{equation}
Based on the calculated total column densities and abundance ratio between $^{13}$C and $^{12}$C in the interstellar medium we can derive the total column density of $^{12}$CO. Then, using the abundance ratios $H_2/^{12}CO=1.1\times 10^4$ \citep{1982ApJ...262..590F} obtained by measuring the extinction of neighboring molecular clouds, we can derive the total column density of H$_2$, 
\begin{equation}
    N(H_2)=[\frac{^{12}CO}{^{13}CO}][\frac{H_2}{^{12}CO}]N(^{13}CO)
    \label{eqA4}
\end{equation}
where $^{12}CO/^{13}CO=6.21\times d_{gc} + 18.71$ and $d_{gc}$ is the distance of the object from the Galactic center\citep{1994ARA&A..32..191W}.
With the obtained H$_2$ column density, we can derive the total gas mass by integrating the column densities over the entire area, i.e., 
\begin{equation}
    M=\mu m_HD^2\int N(H_2)d\Omega
    \label{eqA5}
\end{equation}
where $\mu$ is the average molecular weight 2.8, $m_H$ is the mass of the hydrogen atom, D is the distance of the object from the Earth, and $\Omega$ is the solid angle occupied by the object on the celestial sphere.
According to the definition in \citet{1992ApJ...395..140B}, the virial parameter is
\begin{equation}
    \alpha=\frac{2E_{kin}}{E_{pol}}=\frac{5\delta^2_vR}{GM}
    \label{eqA6}
\end{equation}
where $\delta_v$ is the velocity dispersion in the direction of sight, G is the gravitational constant, R and M are the radius and mass of a cloud, respectively.
From the equation~(\ref{eqA6}), the virial mass can be expressed as 
\begin{equation}
    M_{vir}=\frac{5\delta^2_vR}{G}\approx209(\frac{R}{pc})(\frac{\Delta v}{km s^{-1}})^2M_{\odot},
    \label{eqA7}
\end{equation}
where $\delta v =2\sqrt{\ln 2}\sigma_v$, and $\sigma_v$ is the intensity-weighted velocity dispersion given by the Dendrogram algorithm.


\bibliography{Maddalena_APJ_R1_submit}{}
\bibliographystyle{aasjournal}



\end{document}